\documentclass[twocolumn,trackchanges]{aastex62}

\usepackage{multirow,booktabs}
\usepackage{xcolor}
\graphicspath{{./}{figures/}}

\received{2019 February 4}
\revised{2019 July 22}
\accepted{2019 July 24}
\submitjournal{ApJ}

\shorttitle{Magnetic field measurements: BP Tau and V347 Aur}
\shortauthors{Flores et al.}

\begin{document}

\title{\Large Measuring the magnetic field of young stars using iSHELL observations: \\ BP Tau and V347 Aur}

\email{caflores@hawaii.edu, msc@ifa.hawaii.edu}
\email{reipurth@hawaii.edu, aboogert@hawaii.edu }

\author{C. Flores}
\author{M. S. Connelley}
\author{B. Reipurth}
\affil{
Institute for Astronomy, University of Hawaii at Manoa, 640 N. Aohoku Place, Hilo, HI 96720, USA \\
}
\author{A. Boogert}
\affil{Institute for Astronomy, University of Hawaii at Manoa, 2680 Woodlawn Drive, Honolulu, HI 96822, USA}
\nocollaboration


\begin{abstract}
While it has been suggested that there is a connection between the magnetic properties and the internal structure of young stars, there have not been enough magnetic measurements to firmly establish such a correlation at the earliest ages. Here, we contribute to this endeavor by presenting stellar parameters and magnetic field strength measurements of BP Tau and V347 Aur, both stars observed with the near-infrared spectrograph iSHELL. We first test the accuracy of our method by fitting synthetic stellar spectra to a sample of nine main and post-main-sequence stars. We report uncertainties of $\sigma_{\rm Teff}$ = 91 K in temperature and $\sigma_{\rm log(g)}$ = 0.14 in gravity. We then apply the modeling technique to BP Tau and measure a surface magnetic field strength of $\langle \rm B \rangle $ =  2.5$^{+0.15}_{-0.16}$ kG, confirming literature results. For this star, however, we obtain a much lower temperature value than previous optical studies ($\Delta \rm T \sim 400$ K) and interpret this significant temperature difference as due to the relatively higher impact of starspots at near-infrared wavelengths than at optical wavelengths. We further apply this technique to the class I protostellar source V347 Aur and measure for the first time its magnetic field strength $\langle \rm B \rangle = $ 1.36$^{+0.06}_{-0.05}$ kG and its surface gravity log(g) = 3.25$^{+0.14}_{-0.14}$. Lastly, we combine our measurements with pre-main-sequence stellar evolutionary models and illustrate the effects produced by starspots on the retrieved masses and ages of young stars.
\end{abstract}



\section{Introduction} 

\label{sec:intro}
Magnetic fields are thought to be one of the fundamental drivers of the evolution of pre-main-sequence stars, playing a very important role in the star--disk interaction. Magnetic fields regulate the angular momentum of the system \citep{White}, prevent the star from reaching disrupting velocities \citep{Bouvier2007}, collimate winds and outflows into jets \citep{Romanova2015}, and channel material from the disk's inner edge onto the star along magnetic field lines \citep{Shu1994,Hussain}. Later in the stellar life, magnetic fields slow the rotation of stars through magnetized winds \citep{Schatzman1962} and are also the precursors of stellar flares and coronal mass ejections. 

The Zeeman effect is the basis for calculating magnetic fields in stars \citep{Zeeman}. Atoms and molecules subject to magnetic fields display a split of their energy levels, which cause the observed spectral lines to also be split. Spectral lines symmetrically distributed around the wavelength where a nonmagnetic line would have formed are called $\pi$ components; lines formed at either side of this unperturbed wavelength are called $\sigma$ components. The Zeeman effect relates the wavelength displacement of the Zeeman components ($\Delta \lambda_B$) to the reference wavelength ($\lambda_{\rm ref}$), the transition's magnetic sensitivity ($\overline{g}$), and the applied magnetic field strength (B)

\begin{equation}
\label{eq:zeeman_broad}
\Delta \lambda_B = 4.67\times 10^{-6} \lambda_{\rm ref}^2 \: \overline{g}  \: \rm B
\end{equation}

in this equation $\lambda$ is measured in microns, the effective Land\'e-g factor $\overline{g}$ is a dimensionless constant, and the magnetic field strength B is measured in kilogauss.

Given the importance of magnetic fields in stars, several magnetic radiative transfer codes have been developed, e.g., MoogStokes \citep{Deen}, COSSAM \citep{Stift2012}, Zeeman2 \citep{Landstreet1988}, and Synthmag \citep{Piskunov1999}, and some high-resolution NIR spectrographs, such as CSHELL \citep{Tokunaga1990,Greene1993} and PHOENIX \citep{Hinkle2003}, have been extensively used to acquire such magnetic observations
\citep[e.g.,][]{Johns-Krull1999,Leone2003,Johns-Krull2007,Yang2008,Lavail}. The small wavelength coverage of these early spectrographs, however, has been a strong limitation ($\sim$0.005 \micron \, for CSHELL and $\sim$0.01 \micron \, for PHOENIX), because only a small number of spectral lines could be acquired in a single observation. This led to the first magnetic field strength measurements of young stars to be performed by modeling only a few NIR lines, and by adopting temperature and gravity measurements from optical observations \citep{Johns-Krull1999,Yang2005,Yang2008,Johns-Krull2007,Yang2011}.

With the development of larger IR detectors, broader bandwidth instruments have been constructed, e.g., iSHELL \citep{Rayner2016}, IGRINS \citep{Park2014}, and SPIROU \citep{Moutou2015}. These new generation instruments allow access to more photospheric lines in a single setting and enable simultaneous and consistent stellar parameter measurements. For instance, \cite{Sokal2018} used IGRINS K-band data to determine the stellar parameters of the well-studied TW Hya source and found their temperature and gravity measurements to differ from optical spectroscopic measurements by $\sim 300 \, \rm K$ and $\sim0.6$ dex, respectively. These differences pose questions as to whether stellar parameters derived for young stars depend on the wavelength region they were observed, whether these parameters depend on the technique used to obtain them, or whether they are intrinsically variable and change over time. 

In the present study, we take advantage of iSHELL's large spectral coverage ($\sim$0.29 \micron \, in the K2 mode\footnote{iSHELL's K2 mode covers the 2.09 $\micron$ to 2.38 $\micron$ range}) and high spectral resolution to study the surface magnetic field strength and the atmospheric stellar properties of two young stars, the class II source BP Tau, and the class I source V347 Aur. We selected BP Tau because this class II source has been extensively studied in the literature \citep{Hartigan1995,Schiavon1995,Johns-Krull1999,Grankin2016} and therefore serves as a benchmark to test the magnetic field models applied to a young star. The class I source V347 Aur, on the other hand, does not have a previous magnetic field measurement in the literature. It was selected because it is a bright source in the infrared (K=8.06) and it displayed strong photospheric lines in the survey conducted by \cite{Connelley}. The observations and data reduction of the two young stars and other nine standard stars plus the Sun are described in Section \ref{sec:observations}. In Section \ref{sec:modeling}, we describe the codes we use and the assumptions we make to model the data. In Section \ref{section:standard_stars}, we model the standard stars to quantify the empirical uncertainties in the derived atmospheric stellar parameters T$_{\rm eff}$, log(g), as well as to gauge our magnetic field detection limit. We present the results for BP Tau and V347 Aur in Section \ref{sec:magnetic-models}. We discuss our results in Section \ref{sec:discussion} and summarize our findings in Section \ref{sec:Summary}.

\section{Observations and Data reduction}
\label{sec:observations}
\subsection{Spectroscopic Observations}
We carried out our observations with the IRTF 3.0m telescope on Maunakea, Hawaii. We used the high-resolution, near-infrared echelle spectrograph iSHELL \citep{Rayner2016} to observe two young stellar objects (YSOs), nine main- and post-main-sequence stars, and the Sun in reflected light from the asteroid Ceres. All sources were observed in good weather conditions with seeing varying from $0\farcs4$ to 2$\arcsec$ and airmasses between 1.0 and 1.6. We observed all our sources in the K2 mode of iSHELL, i.e., from 2.09 to 2.38 $\micron$, with the $0\farcs75$ slit to achieve a spectral resolution of R = $47\,000$. Immediately after we observed each one of our science targets, we acquired quartz lamp spectra for flat-fielding, thorium-argon lamp spectra for wavelength calibration, and an A0 standard star for telluric correction. At the end of each observing night, we obtained dark frames that matched the integration times of our science targets and telluric standard stars.  Tables \ref{table:Observation_summary} and \ref{table:Observation_log} summarize information about the observations. 

\begin{deluxetable*}{lcccc}
\tabletypesize{\scriptsize}
\tablecaption{Observation Summary \label{table:Observation_summary}}
\tablewidth{0pt}
\tablecolumns{5}
\tablehead{
\colhead{Name} &  \colhead{Int. Time (minutes)} & \colhead{S/N\tablenotemark{a}} & \colhead{Telluric Std.} & \colhead{Obs. Date}}
\startdata
BP Tau & 10 & 163 & HD 47596 & 2017 Nov 6 \\
V347 Aur & 10 & 138 & HD 29573 & 2018 Aug 31 \\
14 Her & 5 &  318 & HD 124683 & 2017 May 17 \\
GJ 411 & 1 &  311 & 23 LMi & 2017 May 17 \\
TYC 1293-2421-1 & 10 & 218 & k Tau & 2017 Nov 6 \\
BD+004988 & 5 &  188 & HD 209932 & 2017 Nov 15 \\
GJ 380 & 10 &  124 & 26 Uma & 2018 Jan 11 \\
GJ 412A & 1 &  133 & HD 99966 & 2018 Jan 12 \\
GJ 436 & 4 &  200 & HD 107655 & 2018 Jan 12 \\
GJ 526 & 1 &  219 & HD 89572 & 2018 Jan 12 \\
Sun (Ceres) & 6 & 330 & 23 LMi & 2018 Jan 13 \\
EPIC 211304446 & 25 &  232 & 23 LMi & 2018 Jan 13 \\
\hline
\enddata
\tablenotetext{a}{Median signal-to-noise ratio calculated from \texttt{Spextool v5.0.2}.}
\end{deluxetable*}

\begin{deluxetable*}{lcccccc}
\tabletypesize{\scriptsize}
\tablecaption{YSOs and Standard Stars \label{table:Observation_log}}
\tablewidth{0pt}
\tablecolumns{7}
\tablehead{
\colhead{Name} & \colhead{Alt. Name} & \colhead{$\alpha$(J2000) } & \colhead{$\delta$(J2000)} & \colhead{J} & \colhead{H} & \colhead{K}}
\startdata
BP Tau & HD 281934 & 04h 19m 15.8s & +29$^{\circ}$ 06$\arcmin$ 26$\arcsec$ & 9.10 & 8.22  & 7.73\\
V347 Aur & IRAS04530+5126 & 04h 56m 57.0s & +51$^{\circ}$ 30$\arcmin$ 50$\arcsec$ &9.99 & 8.82 & 8.06 \\
14 Her & GJ 614 & 16h 10m 24.3s & +43$^{\circ}$ 49$\arcmin$ 03$\arcsec$ & 5.15 & 4.80 & 4.71 \\
GJ 411 & HD 95735 & 11h 03m 20.1s & +35$^{\circ}$ 58$\arcmin$ 11$\arcsec$ & 4.20 & 3.64 & 3.34 \\
TYC 1293-2421-1 & & 05h 02m 42.9s & +20$^{\circ}$ 50$\arcmin$ 38$\arcsec$ & 8.39 & 7.88 & 7.68 \\
BD+004988 & TYC 577-1200-1 & 23h 19m 22.7s & +01$^{\circ}$ 42$\arcmin$ 29$\arcsec$ & 8.11 & 7.59 & 7.43 \\
GJ 380 & HD 88230 & 10h 11m 22.1s & +49$^{\circ}$ 27$\arcmin$ 15$\arcsec$ & 3.97 & 3.27 & 3.26 \\
GJ 412A & & 11h 05m 28.5s & +43$^{\circ}$ 31$\arcmin$ 36$\arcsec$ & 5.53 & 5.00 & 4.76 \\
GJ 436 & & 11h 42m 11.0s & +26$^{\circ}$ 42$\arcmin$ 23$\arcsec$ & 6.90 & 6.31 & 6.07 \\
GJ 526 & HD 119850 & 13h 45m 43.7s & +14$^{\circ}$ 53$\arcmin$ 29$\arcsec$ & 5.18 & 4.78 & 4.41 \\
EPIC211304446 & TYC 811-1076-1 & 08h 54m 06.1s & +09$^{\circ}$ 56$\arcmin$ 19$\arcsec$ & 8.64 & 8.24 & 8.10 \\
\enddata
\tablecomments{Coordinates and photometry are from SIMBAD.}
\end{deluxetable*}

\subsection{Data Reduction}
\label{sec:datareduction}
We reduced the data using a new version of \texttt{Spextool} \citep{Cushing2004},  v5.0.2, which was specifically designed to reduce iSHELL data\footnote{\url{http://irtfweb.ifa.hawaii.edu/research/dr_resources/}}. Using the \texttt{xspextool} task, we first created the normalized flat-field images and wavelength calibration files. Then, we extracted the spectra of each star (science and telluric standards) using the point source extraction configuration. To increase the signal-to-noise ratio (S/N) of the final spectra, we used \texttt{xcombspec} to combine the individual multiorder stellar spectra and \texttt{xmergeorders} to merge multiorder spectra into a single continuous spectrum. We used \texttt{xtellcor} \citep{vacca2003} to remove atmospheric absorption lines from the science spectra. Then, we used \texttt{xcleanspec} to eliminate any deviant or negative pixels caused by imperfections in the infrared array. We smoothed the R = $47\,000$ spectra with a nine-pixel-wide Savitzky-Golay function in \texttt{xcleanspec}, as this function smooths the input spectrum while preserving the average resolving power. Finally, we normalized the continuum of the reduced spectra of each source using a first-order broken power-law function with the  \texttt{specnorm}\footnote{\url{ftp://ftp.ster.kuleuven.be/dist/pierre/Mike/IvSPythonDoc/plotting/specnorm.html}} code.

\section{Modeling}
\label{sec:modeling}
Stellar synthetic spectral codes have been widely used by numerous authors to derive stellar parameters of young stars \citep[e.g.,][]{Valenti1995,Doppmann,Johns-Krull2007}. Modeling the strength and shape of stellar absorption lines allows us to derive fundamental properties of young sources such as the $\rm T_{eff}$, $\rm \log(g)$, and $v\sin(i)$. The computation of such synthetic spectra can be generally divided into four main parts: the radiative transfer code, the stellar atmospheric models, the spectrograph response, and the list of atomic and molecular transitions. The correct interplay among all the parts is crucial to obtain accurate and reliable synthetic models. In this section, we describe the codes we used and the assumptions we made to compute the synthetic spectra.

\subsection{Model Description}
\label{subsec:synthetic-spectrum-models}
We use MoogStokes \citep{Deen}, a plane-parallel  Local Thermodynamic Equilibrium (LTE) magnetic radiative transfer code to synthesize stellar spectra. This code is a modified version of the classical spectral synthesis code \texttt{MOOG} \citep{Sneden}, with the modifications needed to include the Stokes terms in the radiative transfer equation, as well as the magnetic effects in the opacity calculation of the stellar atmosphere. MoogStokes assumes a radial and uniform magnetic field in the radiative transfer computation, while in the nonmagnetic case (i.e., when $\langle \rm B \rangle $ = 0 kG) the MoogStokes calculations coincide with the results of \texttt{MOOG}.

MoogStokes requires a stellar atmospheric model as an input into the calculation of the radiative transfer equation. We use the \texttt{MARCS} models \citep{Gustafsson}, which are 1D hydrostatic LTE plane-parallel and spherical stellar atmospheric models suitable for stars with $\rm T_{eff}$ between $2\,500$ K and $8\,500$ K; log(g) from 0.0 to 5.5 (cm s$^{-2}$) in log scale\footnote{Throughout the text, we will always refer to gravity measurements in cgs units and log scale.}; [M/H] from -2 to 2 dex compared to solar, and microturbulence ($v_{\rm micro}$) from 0 to 5 km s$^{-1}$. Although the \texttt{MARCS} atmospheric models allow for anomalous atmospheric abundances, such as the alpha enhancement of elements and CN-cycled abundances, we adopted, for simplicity, a solar composition in all our models.

We empirically measured iSHELL's spectral profile and found that the spectral profile using the 0\farcs75 slit width in the K2 mode is well characterized by the convolution of a boxcar function and a Voigt profile. In Appendix \ref{sec:appendix_iSHELL_PSF} we describe our procedure to obtain the empirical spectral profile of iSHELL and provide an analytical description. 

Line transition parameters are quantities that influence the shape and strength of atomic and molecular spectral lines. These parameters, such as the reference wavelength $\lambda_{ref}$, the oscillator strength log(gf), and the van der Waals constants VdW, are often not precise enough to perform high-precision spectroscopic studies of stars \citep{Shetrone2015,Andreasen,Lavail}. Therefore, we initially adopted atomic line parameters from the VALD3\footnote{\url{http://vald.astro.uu.se/}} database \citep{Ryabchikova2015} and CO molecular line transitions from the HITEMP\footnote{\url{https://www.cfa.harvard.edu/hitran/HITEMP.html}} database \citep{Rothman2010}, but then implemented modifications to these values to improve the precision of the calculations. In Appendix \ref{subsec:linelist-modification}, we explain how we modified the line transition parameters of a number of atomic lines by comparing MoogStokes models to solar observations.

\subsection{Diagnostic Spectral Lines} \label{subsec:diagnostic_lines}
iSHELL's large spectral coverage ($\sim0.29$ \micron $\,$  in the K2 mode) enables us to access many stellar photospheric lines in a single observation. Ideally, we would like to use all the information contained in this spectral range to compare our models to the data. In practice, however, it is too computationally expensive for us to perform such modeling over the full $\sim 0.29$ \micron $\,$ bandwidth. Therefore, we have instead selected specific wavelength regions with prominent photospheric absorption lines (in the spectrum of low-mass stars) that contain enough spectral information to allow us to derive the atmospheric stellar parameters we are interested in. Selecting strong photospheric lines is important in this modeling framework, as these lines can be used to derive stellar parameters of young stars, even when the circumstellar emission significantly veils the spectrum.

\begin{figure*}
\epsscale{1.0}
\plotone{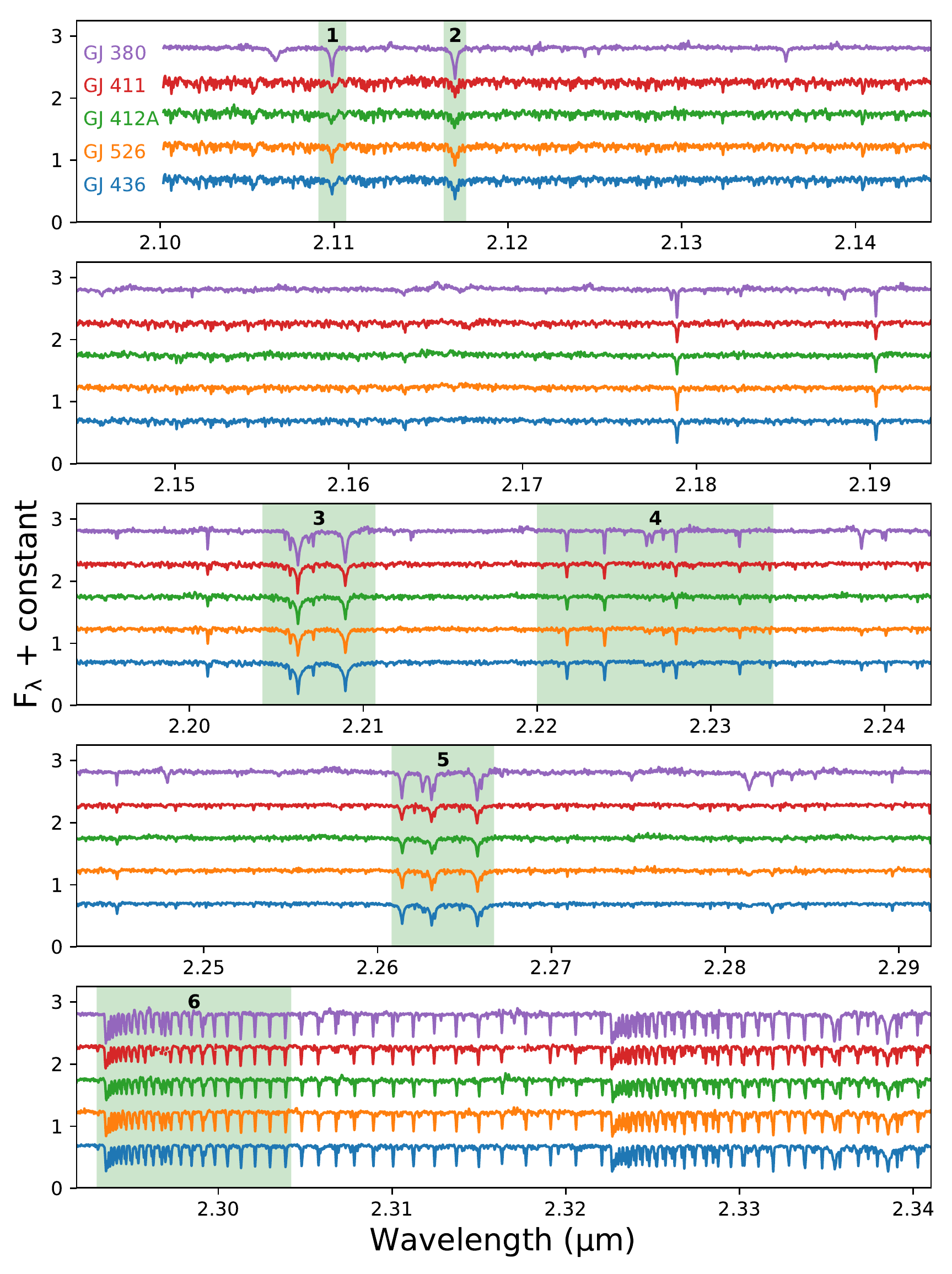}
\caption{Sample of M dwarfs in order of descending temperature. The shaded green regions illustrate the six wavelength ranges we used to fit our models to the data. GJ 380 $\rm T_{eff}= 3\,970\, K$, GJ 411 $\rm T_{eff}= 3\,604\, K$, GJ 412A $\rm T_{eff}= 3\,579\, K$, GJ 526 $\rm T_{eff}= 3\,555\, K$, GJ 436$\rm T_{eff}= 3\,401\, K$ (see Table \ref{table:Standard_star_params})}. \label{fig:spectroscopic_dwarfs}
\end{figure*}

\begin{figure*}
\epsscale{1.0}
\plotone{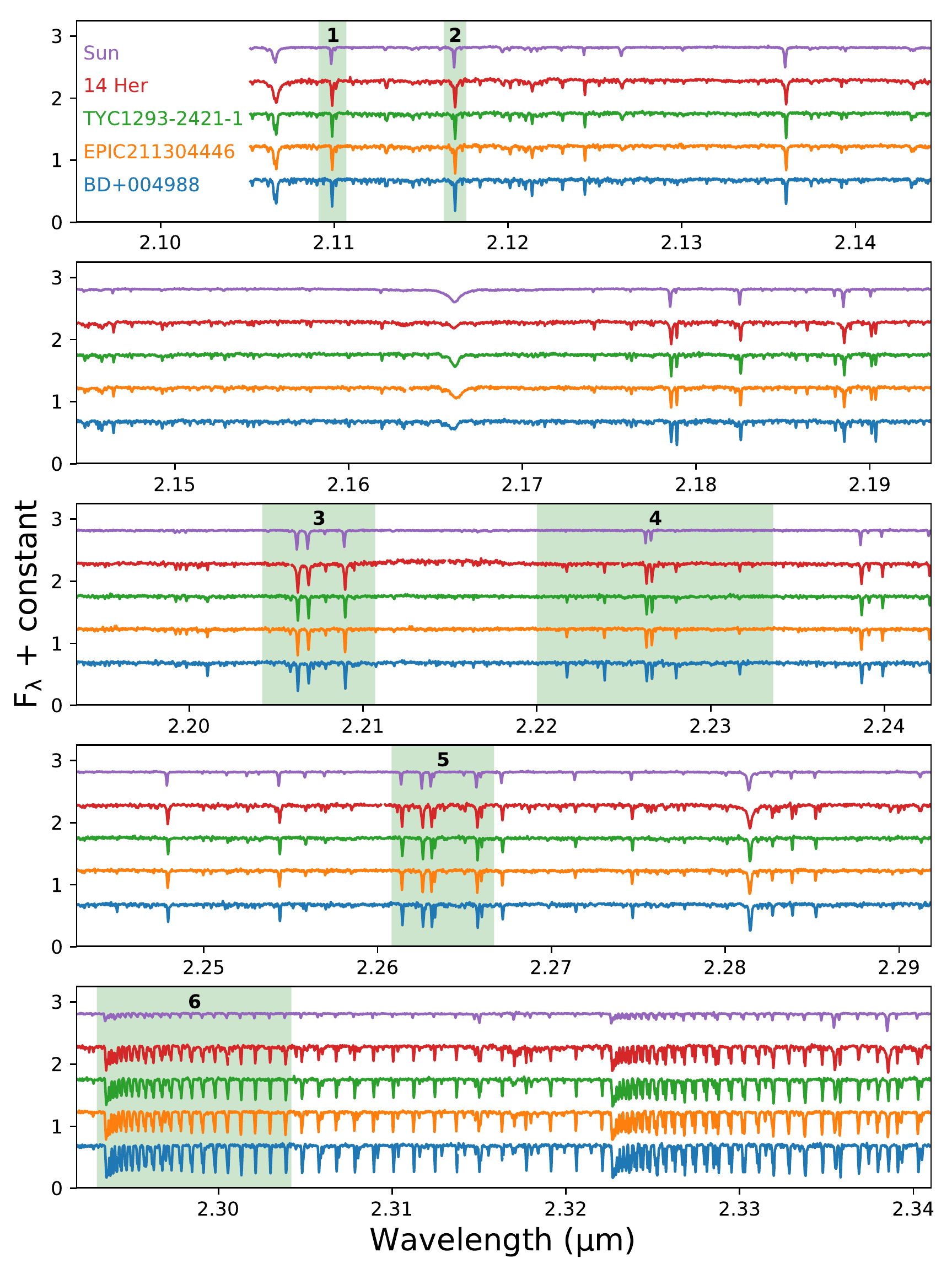}
\caption{Sample of main- and post-main-sequence stars in order of descending temperature. 14 Her $\rm T_{eff}= 5\,430\, K$, TYC1293-2421-1  $\rm T_{eff}= 5\,059\, K$, EPIC211304446 $\rm T_{eff}= 4\,856\, K$, BD+004988  $\rm T_{eff}= 4\,500\, K$.} \label{fig:spectroscopic_giants}
\end{figure*}

We selected a combination of atomic and molecular lines, as some of them are highly sensitive to changes in the temperature and gravity of stars, e.g., the Na lines at 2.2062 \micron \, and 2.2089 \micron \,, and the Ca lines at 2.2614 and 2.2657 \micron. Others have large sensitivity to magnetic fields (large Land\'e-g factors), such as the Ti lines at 2.2217 and 2.2316 \micron \,, and molecular features\footnote{Water lines in the K band can be used to estimate the temperature and gravity of cool stars \citep{Cushing2005}. However, we chose not include these lines in our modeling analysis as their effect on continuum-normalized spectra is rather small, reaching depths of $\sim10\%$.} such as the CO (2-0) rovibrational transitions, starting at 2.2935 \micron \,, which are useful to constrain the rotational velocity of the stars, due to their low magnetic sensitivity. We divided our diagnostic lines into six spectral regions (see Figures \ref{fig:spectroscopic_dwarfs} and \ref{fig:spectroscopic_giants}) and summarize them in Table \ref{table:wavelength_regions}. These six regions provide an approximate spectral coverage of 0.038 \micron \, with more than 25 photospheric lines as parameter diagnostics.

\begin{deluxetable}{ccc}
\tablecaption{Spectral Regions Used in the Computation of the Synthetic Models \label{table:wavelength_regions}}
\tablehead{
\colhead{Region number} & \colhead{Spectral range}  & \colhead{Main spectral} \\ 
  & (\micron)  & lines }
\startdata
 1 & 2.1091--2.1107 & Al, Fe  \\
 2 & 2.1163--2.1176 & Al, Fe, Ni \\
 3 & 2.2055--2.2100 & Na, Sc, Si \\
 4 & 2.2200--2.2336 &  Ti, Fe, Sc \\
 5 & 2.2608--2.2664 & Ca, Fe, Ti, S \\
 6 & 2.2930--2.3042 & CO(2-0) \\
\enddata
\tablecomments{Wavelength values are defined in vacuum.}
\end{deluxetable}

\subsection{Modeling the Data}
\label{subsect:how_we_fit_models}
To extract stellar parameters from the data, we compared our observed stellar spectra to synthetic spectra generated with the codes described in Section \ref{subsec:synthetic-spectrum-models}. We started the modeling procedure by performing a radial velocity (RV) correction to the data. We did this by comparing, for each of the wavelength regions defined in Section \ref{subsec:diagnostic_lines}, the observed spectrum to a MoogStokes model with approximate literature stellar parameters. The RV correction was performed at a subpixel level, where 1 pixel corresponds to 1 km s$^{-1}$. The uncertainty in the RV correction is dominated by the wavelength solution obtained from \texttt{Spextool v5.0.2}, which is typically of the order of half a pixel or 0.5 km s$^{-1}$. Then, to find the synthetic model that better reproduces an observed spectrum, we implemented a Markov chain Monte Carlo (MCMC) parameter-space search using the \texttt{emcee}\footnote{\url{http://dan.iel.fm/emcee/current/}} code \citep{Foreman-Mackey}.

The MCMC algorithm is a Bayesian inference method used to obtain an unknown probability distribution function from a sequence of random samples. This computed probability distribution allows us to obtain the stellar parameters of interest and estimate their formal errors  directly from the posterior probability distribution function \citep{Andrae}. We defined the most likely values of the stellar parameters as the median of the posterior distribution after the MCMC run has converged; likewise, we assessed the formal uncertainty for every parameter as the 16th and 84th percentiles of the distributions.

In our modeling approach, we simultaneously fitted seven out of eight possible stellar parameters for each star in our sample. The parameters we allowed to vary were temperature, gravity, surface magnetic field strength, rotational velocity, microturbulence velocity, metallicity, IR K-band veiling, and CO abundance. The T$_{\rm eff}$, log(g), $\langle \rm B \rangle $, $v_{\rm micro}$, and [M/H] are stellar parameters directly implemented into the radiative transfer code. To include the $v\sin(i)$ effects in the models, MoogStokes convolves the output spectra with a rotational kernel following the disk integration method developed by \cite{Valenti1996}. When we analyzed the spectrum of young stars, we included the IR K-band veiling parameter (r$_K$), as it accounts for the excess of infrared radiation emitted from the circumstellar material. The infrared K-band veiling, formally defined as r$_K \equiv$ F$_{\rm K,ex}$/F$_{\rm K*}$ (where F$_{\rm K,ex}$ is the K-band flux from the circumstellar environment and F$_{\rm K*}$ is the stellar photospheric K-band flux) affects the depth of the photospheric lines in young stars, making them appear weaker if the IR excess is large. We defined the CO abundance parameter as a parameterization of all the processes that affect the formation and destruction of CO on the surface of young and evolved stars. We included this parameter because the assumption of a single `solar' CO abundance for all the standard stars largely disagreed with our observations. A CO abundance value equal to 1 corresponds to solar CO abundance, while lower or higher values correspond to a CO deficiency or enhancement with respect to solar, respectively.

In the MCMC routine, we set our priors to match the full range of stellar parameters of the \texttt{MARCS} atmospheric models (see Section \ref{subsec:synthetic-spectrum-models}), except for the metallicity parameter where we restricted it to be within -1.0 and 1.0 dex compared to solar, as none of our stars are extremely metal enriched or metal deficient. We allowed the projected rotational velocity parameter to vary between 3 and 20 km s$^{-1}$. For the IR K-band veiling parameter, we adopted a prior that allows it to vary between 0 (no veiling) and 5 (where the strength of the lines is 1/6 of the unveiled strength). We allowed the CO abundance parameter to vary between 0.2 and 10. For each star, we ran at least $\sim80\,000$ models, performing linear interpolation in the \texttt{MARCS} stellar atmospheric models when necessary to access portions of the parameter space that are not defined in the grid of atmospheric models.

\begin{figure*}[ht!]
\epsscale{1.0}
\plotone{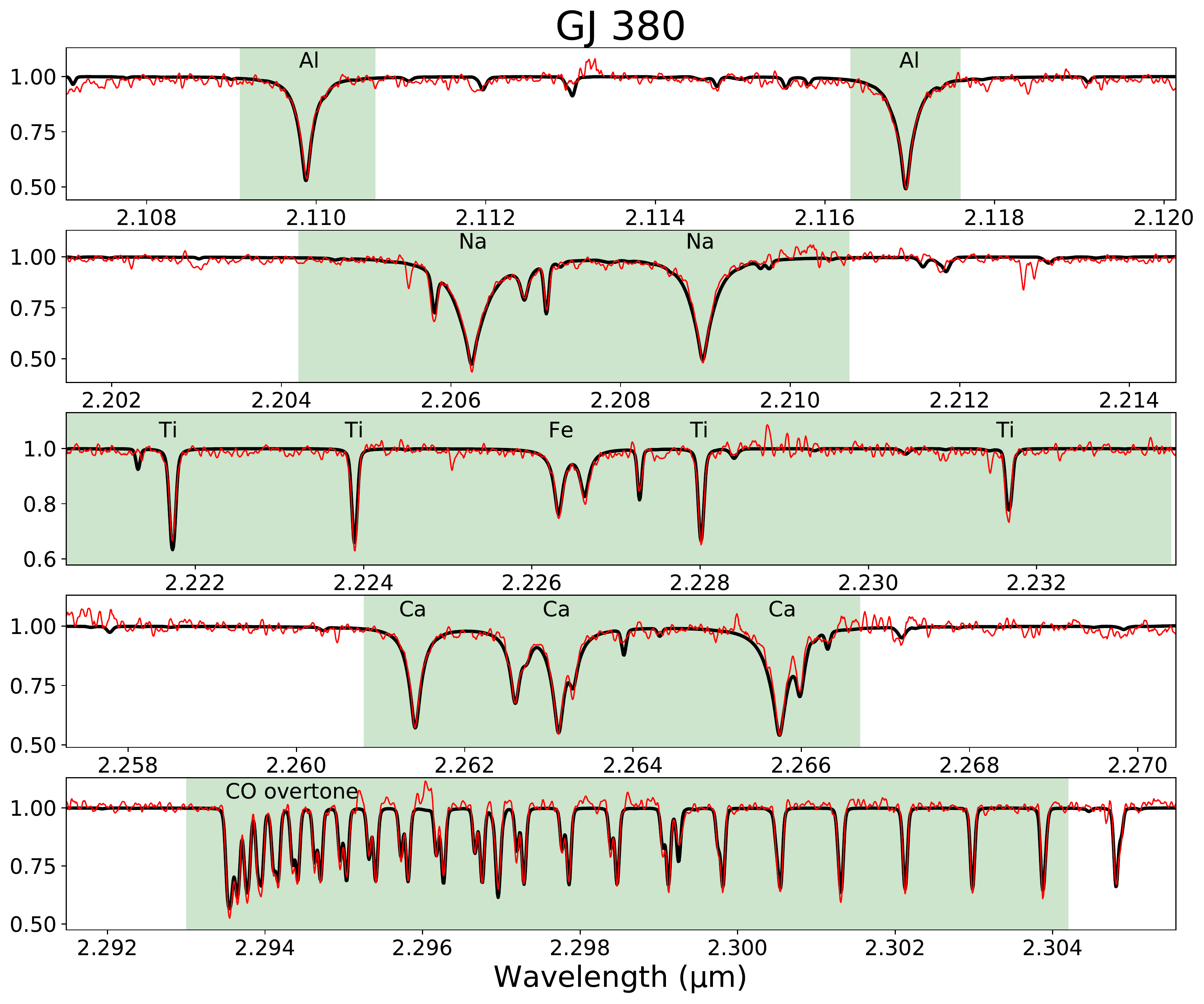}
\caption{Comparison between the observed spectrum of the M-dwarf GJ380 (in red) and the best fit model (in black). All the panels in the figure have the same bandwidth. \label{fig:GJ380}}
\end{figure*}

\begin{figure*}[ht!]
\epsscale{1.0}
\plotone{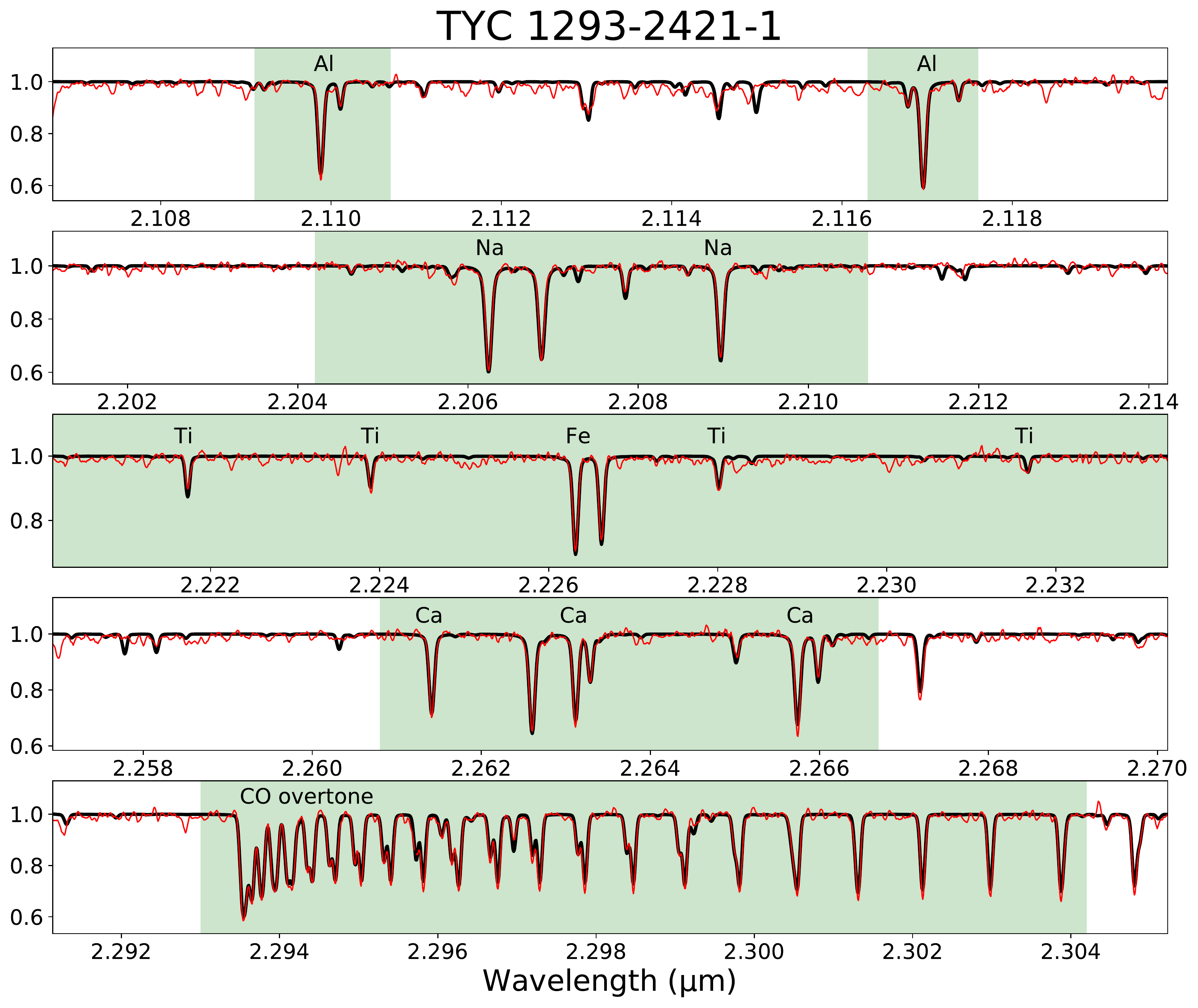}
\caption{Comparison between the observed spectrum of the red giant TYC1293-2421-1 (in red) and the best fit model (in black).\label{fig:TYC1293}}
\end{figure*}

\section{Modeling Main- and Post-main-sequence stars} \label{section:standard_stars}
To validate our model procedure and to understand the real uncertainties in the derived stellar parameters, we modeled nine main- and post-main-sequence standard stars using the synthetic spectrum code described in Section \ref{sec:modeling}. Three of the standard stars in our sample are red giants with accurate gravities calculated from asteroseismology. Five of the stars are M dwarfs with temperatures calculated from interferometric radii measurements; the remaining star is a K dwarf with stellar parameters measured from high-resolution optical spectroscopy. 
We selected our sample of standard stars to span a temperature range from $\sim$3\,400 K to $\sim$5\,300 K and gravities from 2.6 to 4.9, which encompass the temperature and gravity ranges of low-mass young stars \citep{Doppmann,Baraffe,Feiden}. 

The main differences between the evolved stars and the low-mass young stars are their rotational velocities, magnetic field strength values, and the amount of circumstellar material around them. Young stars are known to have a wide range of projected rotational velocities \citep{Covey2005,Bouvier2014}, ranging from a few km s$^{-1}$ to several tens of km s$^{-1}$. Most, if not all, low-mass young stars show kilogauss-strength magnetic fields on their surfaces \citep{Johns-Krull2007,Yang2011}; likewise, young stars typically exhibit a large amount of circumstellar material detected at IR and submillimeter wavelengths \citep{Cieza2005}. On the other hand, most late-type main-sequence and post-main-sequence stars have low rotational velocities \citep[$\leq$ 10 km s$^{-1}$,][]{Jenkins2009,Carlberg2011}, typically very weak surface magnetic field strengths \citep[see][for $\langle\rm |B_V| \rangle$ measurements\footnote{The longitudinal surface magnetic field averaged over the stellar surface.} of main-sequence stars]{Vidotto2014}, and a very small or nondetectable amount of circumstellar material around them. Therefore, in the following analysis, we assumed that main-sequence and post-main-sequence stars have a negligible excess of infrared radiation, and we do not include the IR K-band veiling parameter in the calculations. The stellar parameters we simultaneously fit for the sample of standard stars are T$_{\rm eff}$, log(g),  $\langle \rm B \rangle$ , $v\sin(i)$, $v_{\rm micro}$, [M/H], and CO abundance. 

Figures \ref{fig:GJ380} and \ref{fig:TYC1293} show an example of our modeling technique applied to the M-dwarf GJ 380 and the red giant TYC 1293-2421-1. In both cases, the best-fit model reproduces well the data and our derived stellar parameters agree within a few $\sigma$ with the stellar parameters found in the literature. In Table \ref{table:Standard_star_params}, we list the literature stellar  parameters of the standard stars and also summarize the stellar parameters we recovered using our modeling approach.

\begin{deluxetable*}{ccccccccc}
\tabletypesize{\scriptsize}
\tablecaption{Literature Stellar Parameters vs. Our Derived Parameters for the Sample of Main- and Post-main-sequence Stars. \label{table:Standard_star_params}}
\tablehead{
&\colhead{Star} & \colhead{$\rm T_{\rm eff}$} & \colhead{log(g) } & \colhead{[M/H]} & \colhead{$v\sin(i)$} & \colhead{$v_{\rm micro}$} & \colhead{$\langle \rm B \rangle $} & \colhead{References}\\
&\colhead{Name} & \colhead{(K)} &  & \colhead{} & \colhead{(km s$^{-1}$)} & \colhead{(km s$^{-1}$)} & \colhead{(kG)} & \colhead{}}
\startdata
\multirow{10}{*}{\rotatebox[origin=c]{90}{Literature Values}}
& 14 Her & 5\,300 $\pm$ 90 & 4.27 $\pm$ 0.16 & 0.5 $\pm$ 0.05 & 1.6 & 0.8 $\pm$ 0.12 &   & 1 \\
& GJ 380 & 4\,081 $\pm$ 15 & 4.643 $\pm$ 0.05 & -0.16 $\pm$ ?  & 1.9 & ? & 0.8 $\pm$ 0.3  & 2,3,8,9  \\
& GJ 526 & 3\,618 $\pm$ 31  & 4.784 $\pm$ 0.05 & -0.3 $\pm$ 0.17 & 1.0 & ? & 0.9 $\pm$ 0.3  & 2,3,4,8 \\
& GJ 411 & 3\,465 $\pm$ 17  & 4.845 $\pm$ 0.05 & -0.41 $\pm$ 0.17  & 0.61 & ? & 1.0 $\pm$ 0.3 & 2,3,4,8  \\
& GJ 412A & 3\,497 $\pm$ 39 & 4.843 $\pm$ 0.05 & -0.4 $\pm$ 0.17 & 2.4 & ? &  & 3,4,8 \\
& GJ 436 & 3\,416 $\pm$ 53  & 4.796 $\pm$ 0.05 & 0.04 $\pm$ 0.17 & ? & ? & 1.1 $\pm$ 0.3 & 2,3,4,7 \\
& TYC 1293-2421-1 & 4\,943 $\pm$ 183  & 2.81 $\pm$ 0.03 & -0.22 $\pm$ 0.52 & ? & ? &   & 6  \\
& EPIC 211304446 &  4\,937 $\pm$ 77 & 3.412 $\pm$ 0.18 & -0.07 $\pm$ 0.16  & ? & ? &   & 5  \\
& BD+004988 & 4\,487 $\pm$ 53 & 2.65 $\pm$ 0.03 & 0.19 $\pm$ 0.1  & ? & ? &   & 6  \\
& Sun & 5\,778 & 4.43 & 0.0 & 2.06\tablenotemark{a} & 1.0 & 0.001 & \\
\hline
\multirow{10}{*}{\rotatebox[origin=c]{90}{Our Results}}
& 14 Her & $5\,430^{+61}_{-60}$  & $4.30^{+0.04}_{-0.04}$ & $0.58^{+0.02}_{-0.10}$ & $<4$ & $1.32^{+0.11}_{-0.24}$  & $0.21^{+0.09}_{-0.06}$ &  \\
& GJ 380 & $3\,970^{+9}_{-10}$  & $4.50^{+0.01}_{-0.01}$ & $0.08^{+0.01}_{-0.01}$ & $<4$ & $1.34^{+0.02}_{-0.01}$  & $0.40^{+0.03}_{-0.03}$ & \\
& GJ 526 & $3\,555^{+6}_{-5}$  & $4.50^{+0.01}_{-0.01}$ & $-0.36^{+0.01}_{-0.01}$ & $<4$ & $0.43^{+0.09}_{-0.09}$  & $0.47^{+0.01}_{-0.01}$ & \\
& GJ 411 & $3\,604^{+4}_{-4}$  & $5.00^{+0.01}_{-0.01}$ & $-0.55^{+0.01}_{-0.01}$ & $<4$ & $0.42^{+0.10}_{-0.10}$ & $0.35^{+0.01}_{-0.01}$ & \\
& GJ 412A & $3\,579^{+11}_{-13}$  & $4.97^{+0.02}_{-0.03}$ & $-0.60^{+0.01}_{-0.01}$ & $<4$ & $0.43^{+0.13}_{-0.16}$ & $0.40^{+0.02}_{-0.02}$ & \\
& GJ 436 & $3\,401^{+4}_{-3}$  & $4.74^{+0.01}_{-0.01}$ & $-0.12^{+0.01}_{-0.01}$ & $<4$ & $0.35^{+0.11}_{-0.11}$ & $0.39^{+0.02}_{-0.01}$ &  \\
& TYC 1293-2421-1 & $5\,059^{+14}_{-11}$  & $3.02^{+0.03}_{-0.04}$ & $0.14^{+0.01}_{-0.01}$ & $<4$ & $1.62^{+0.02}_{-0.03}$ & $0.10^{+0.05}_{-0.03}$ & \\
& EPIC 211304446 & $4\,856^{+9}_{-10}$  & $3.36^{+0.02}_{-0.01}$ & $0.07^{+0.01}_{-0.01}$ & $<4$ & $1.38^{+0.05}_{-0.08}$ & $0.31^{+0.02}_{-0.03}$ & \\
& BD+004988 & $4\,500^{+4}_{-4}$  & $2.77^{+0.02}_{-0.01}$ & $0.08^{+0.01}_{-0.01}$ & $<4$ & $1.43^{+0.02}_{-0.03}$ & $0.21^{+0.02}_{-0.02}$  &  \\
& Sun & $5\,865^{+93}_{-128}$ & $4.43^{+0.03}_{-0.02}$ & $0.03^{+0.02}_{-0.02}$ & $<4$ & $0.83^{+0.09}_{-0.11}$ & $0.08^{+0.08}_{-0.06}$ & \\
\enddata
\tablecomments{The uncertainty quoted in the stellar parameters derived in this work corresponds to the 16th and 84th percentiles of the posterior  distributions functions derived from the MCMC runs. Actual errors are significantly larger than the tabulated formal uncertainties in derived model parameters (See Sections \ref{subsubsec:gravity_uncert} and \ref{subsubsec:temp_uncert}). The ``?'' sign reported in Table \ref{table:Standard_star_params} means that we did not find literature values for this parameter or the parameter's uncertainties were not stated. The last column lists the references: 1: \cite{Gonzalez1999}, 2: \cite{Moutou2017}, 3: \cite{Boyajian2012}, 4: \cite{Rojas-Ayala2012}, 5: \cite{Huber2016}, 6: Grunblatt et al. (in prep), 7: \cite{vonBraun2012}, 8: \cite{Houdebine2010}, and 9: \cite{Anderson2011}.\\
}
\tablenotetext{a}{Rotational velocity measured at the solar equator.}
\end{deluxetable*}

\begin{figure}[ht!]
\epsscale{1.0}
\plotone{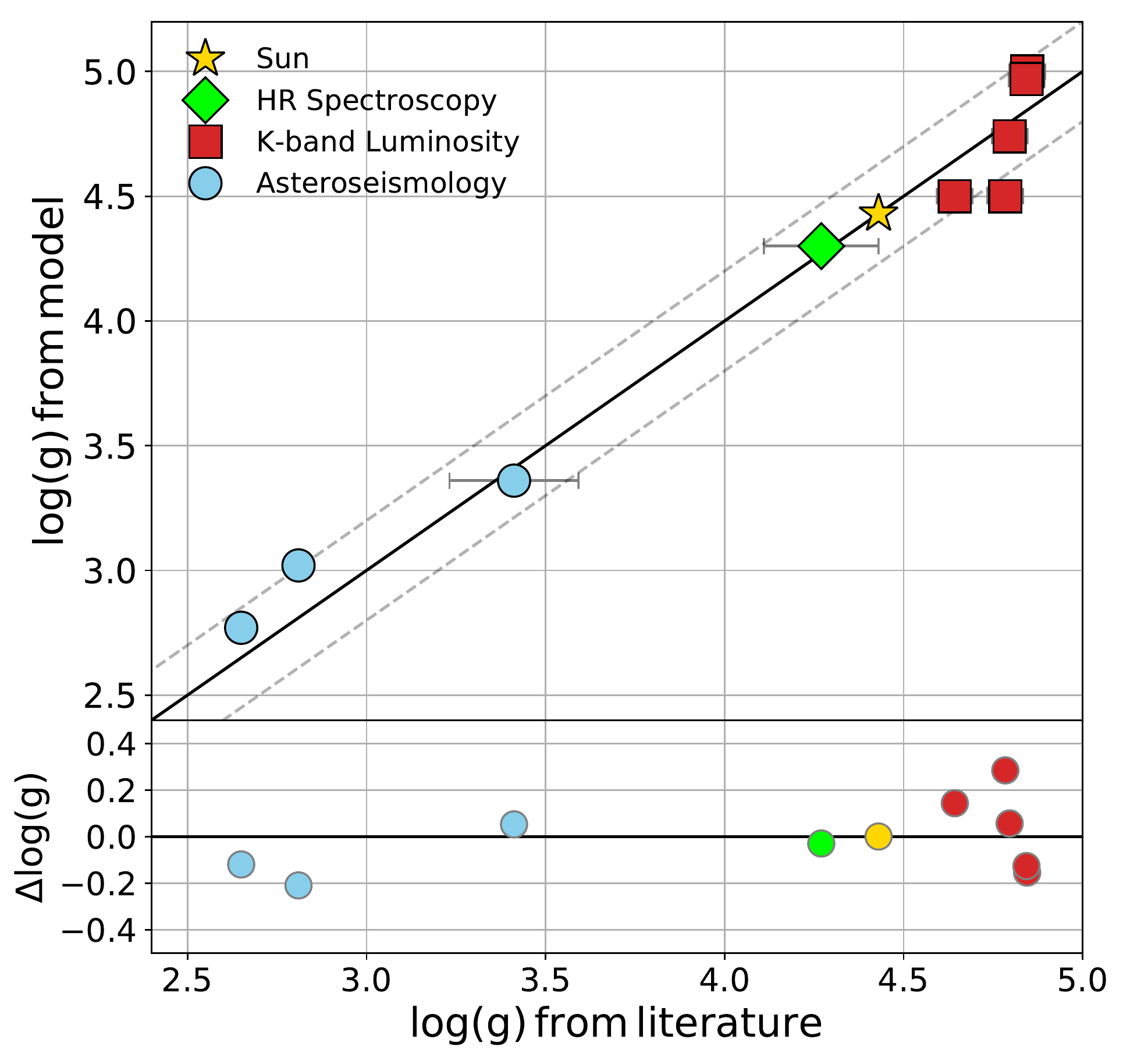}
\caption{Comparison between the literature surface gravity vs. our derived surface gravity for the sample of main- and post-main-sequence stars. The solid black line corresponds to a one-to-one correlation between the model and the literature parameters. The dashed gray lines correspond to differences of $\pm$0.2 dex between both measurements. In this figure, stars with gravities measured from asteroseismology are represented by circles, stars with gravities measured from K-band absolute luminosity and evolutionary models by squares, and a diamond represents the star with gravity measured from high-resolution optical spectroscopy.  \label{fig:Logg_Literature_comparison}}
\end{figure}

\begin{figure}[ht!]
\epsscale{1.0}
\plotone{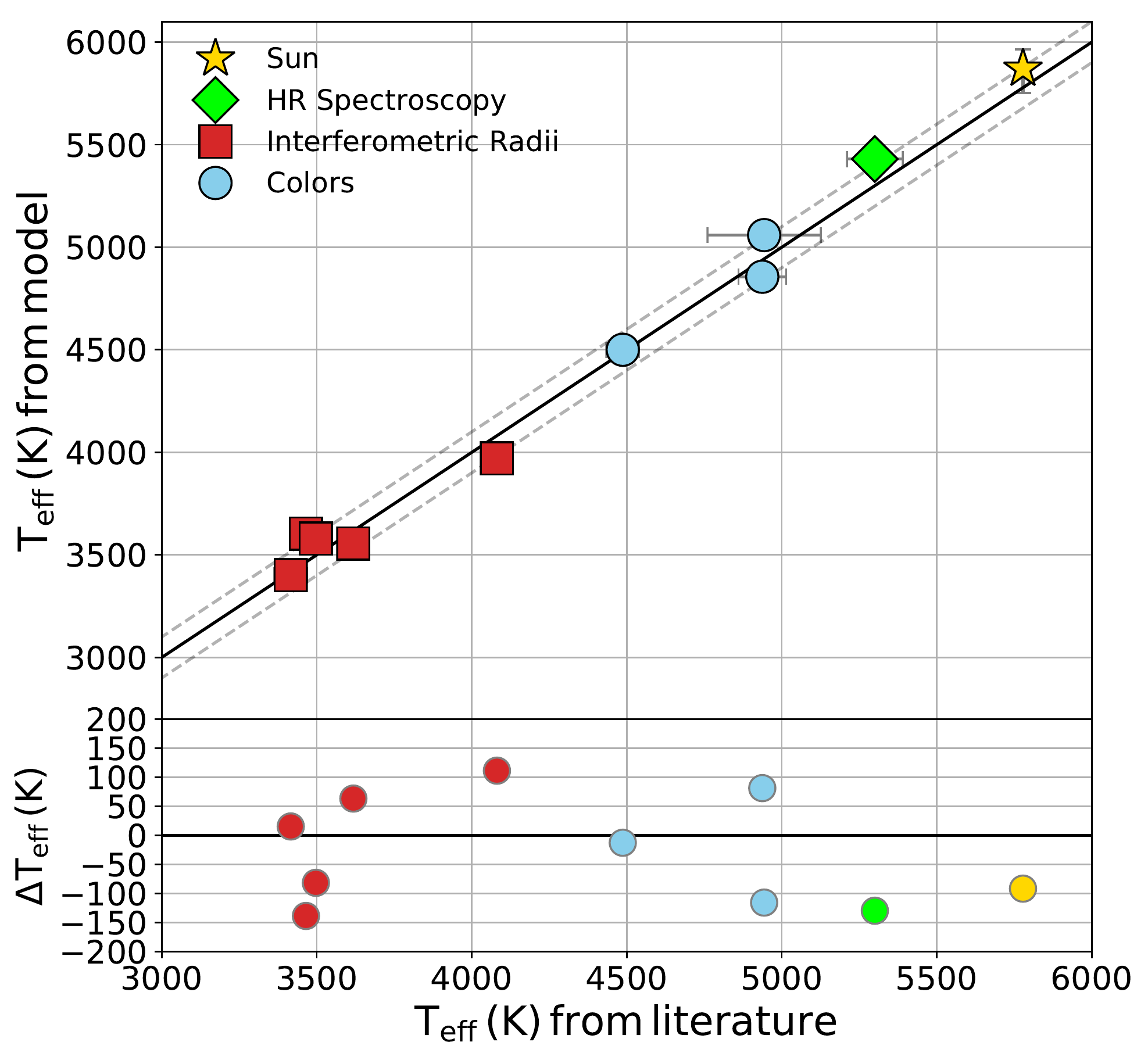}
\caption{Comparison between the literature effective temperatures vs. our derived effective temperatures for the sample of main- and post-main-sequence stars. The solid black line corresponds to a one-to-one correlation between the model and the literature parameters. The dashed gray lines correspond to differences of $\pm$100 K between both measurements. Circles represent stars with temperatures measured from colors and stellar population synthesis models, squares correspond to stars with temperatures obtained from interferometric radii measurements, and a diamond represents the star with temperature measured from high-resolution optical spectroscopy. \label{fig:Teff_Literature_comparison}}
\end{figure}

\begin{figure}[ht!]
\epsscale{1.0}
\plotone{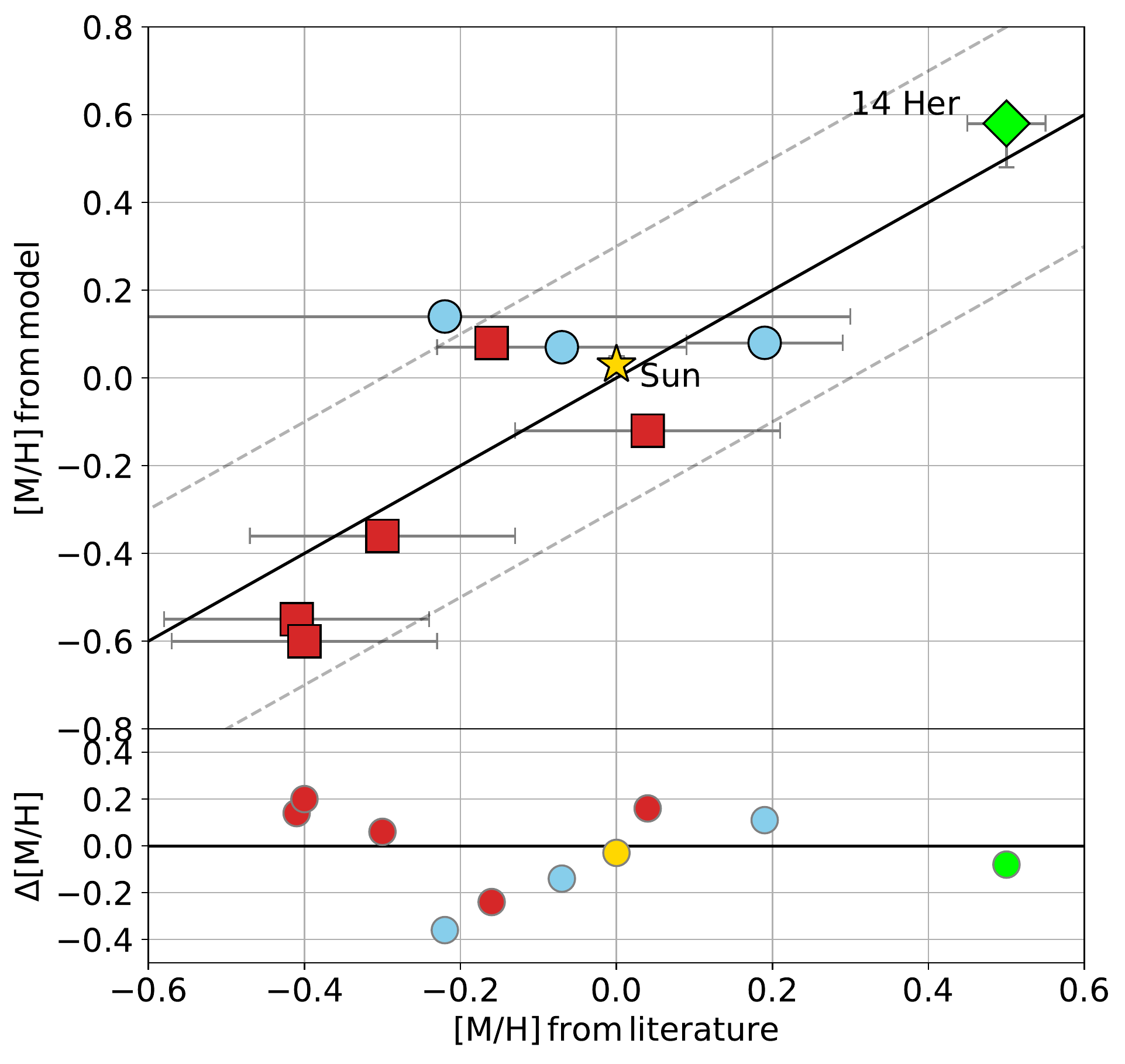}
\caption{Comparison between the literature metallicities [M/H] vs. our derived metallicities for the sample of main- and post-main-sequence stars. The solid black line corresponds to a one-to-one correlation between the model and the literature parameters. The dashed gray lines correspond to differences of $\pm$0.3 dex between both measurements. The symbols used in this plot are the same as those used in Figure \ref{fig:Logg_Literature_comparison} and correspond to gravity measurements. \label{fig:Met_Literature_comparison}}
\end{figure}

\subsection{Gravity Measurements} \label{subsubsec:gravity_uncert}
The first stellar parameter of the sample of standard stars we investigate is the surface gravity. Figure \ref{fig:Logg_Literature_comparison} shows a comparison between the literature log(g) values and the log(g) values derived from our models.  In this figure, dwarf stars with log(g) values calculated from absolute K-band mass-luminosity relations \citep{Boyajian2012} are represented as squares, the dwarf star with log(g) value obtained from high-resolution optical spectroscopy \citep{Gonzalez1999} is shown as a diamond, and the giant stars with accurate gravities calculated from asteroseismology \citep{Huber2016} are shown as circles. The gravity of the Sun is also shown in this plot, and it is represented by an orange star.

The full sample of standard stars can be divided into six dwarf stars and three giant stars. Our derived log(g) values for the six dwarfs and the Sun agree within 0.30 dex with the literature values. Our results for the giant stars, on the other hand, agree within 0.25 dex with the gravities derived from the literature. If we compare our log(g) results with literature results, we measure a mean gravity difference of $\rm \Delta log(g) = 0.02$ and a 1$\sigma$ standard deviation of $\rm \sigma_{log(g)} = 0.14$. Hereafter, we adopt a model uncertainty of $\rm \sigma_{log(g)} = 0.14$ when deriving the surface gravity of YSOs, which will be added in quadrature to the formal uncertainty obtained from the MCMC models.

\subsection{Temperature Measurements}\label{subsubsec:temp_uncert}
We now turn to the temperature measurements of the main- and post-main-sequence stars. Figure \ref{fig:Teff_Literature_comparison} shows how the temperature derived from our work compares to the temperature found in the literature (see Table \ref{table:Standard_star_params}). We find that our derived temperature values agree within 200 K with the literature temperature values. By comparing both measurements, we calculate a mean temperature difference of $ \rm \Delta T_{eff} = -36 \, K$ and a 1$\sigma$ standard deviation of $\rm \sigma_{Teff} = 91 \, K$. Therefore, we consider a model uncertainty of $\rm \sigma_{Teff} = 91\, K$ when deriving effective temperatures of YSOs, which will be added in quadrature to the formal uncertainties obtained from the MCMC models.

\subsection{Metallicity Measurements}
The metallicity parameter is the most uncertain stellar parameter of the sample of standard stars. The mean uncertainty value of the literature metallicity is 0.13 dex. Four of our M-dwarf sources have metallicities derived from low-resolution spectroscopy ($\rm R <3\,000$) while all three red giants have metallicities calculated from population synthesis models \citep{Huber2016}. Only one star in our sample has a high-precision metallicity measurement obtained from high-resolution optical spectroscopy (14 Her). Figure \ref{fig:Met_Literature_comparison} shows the comparison between literature metallicities and our measurements. The metallicity of 14 Her agrees with our measurements within 0.03 dex. Comparing our results with literature measurements, we find a mean metallicity difference of $\rm \Delta [M/H] = 0.03 \,dex$ and a 1$\sigma$ standard deviation of $\rm \sigma_{[M/H]} = 0.14\, dex$. 

\subsection{Rotational Velocities and Microturbulence Values}
Only five of our standard stars have $v\sin(i)$ measurements from the literature (all dwarf stars), and they all rotate slower than 3 km s$^{-1}$. Our velocity resolution is $\sim 6$ km s$^{-1}$ (6 pixels), and we estimate that the lowest rotational velocity we can confidently measure is $\sim$4 km s$^{-1}$. Because all our measurements are below 4 km s$^{-1}$, we can only provide upper limits to the values found in the literature and we cannot draw further conclusions on our sample of standard stars. Additionally, we found only one source with a microturbulence velocity measurement in the literature. \cite{Gonzalez1999} derived a $v_{\rm micro}$ =  0.8 km s$^{-1}$ for 14 Her, which agrees within 2$\sigma$ with our measurement. The mean microturbulence value we derive for the sample of dwarf stars is $v_{\rm micro}$ = 0.7 km s$^{-1}$ with a 1$\sigma$ dispersion of $\Delta v_{\rm micro}$ = 0.4 km s$^{-1}$, while the average micro turbulence value for our asteroseismic giants is 1.42 km s$^{-1}$.  


\subsection{Magnetic field detection limits}
\label{subsection:magnetic_detection_lims}
Our next test focuses on determining the lowest magnetic field strength we can confidently measure using our modeling technique. Four of the M dwarfs from our sample of standard stars have literature magnetic field measurements from optical high-resolution spectroscopy. \cite{Moutou2017} used ESPADONS on CFHT to measure the magnetic field strength of a large sample of M dwarfs by comparing the excess broadening of the magnetically sensitive FeH line at 990.5075 nm against the magnetically insensitive FeH line at 995.0334 nm. In their study, \cite{Moutou2017} compared a small number of their stars against literature magnetic fields from \cite{Shulyak2017} and noticed that for slowly rotating stars their magnetic field values were higher than the magnetic field measurements obtained from Zeeman synthetic modeling \cite[see figure 3 from][]{Moutou2017}. Comparing \cite{Moutou2017} magnetic field results with our measurements, we see that their results are also higher, by a factor of $\sim$2, than our derived values (see Table \ref{table:Standard_star_params}), possibly indicating a systematic error on their method at low rotational velocities. In addition, because the Zeeman effect is more sensitive to magnetic broadening at longer wavelengths (eq. \ref{eq:zeeman_broad}), it is not surprising that our study is capable of detecting weaker magnetic field strengths than optical studies.

Although none of our giant stars have magnetic field measurements, spectropolarimetric studies of single giant stars show that slowly rotating post-main-sequence stars have very weak magnetic fields ($\langle\rm |B_V| \rangle<0.1 \,kG$;) \citep[and references therein]{Linsky2015}. We therefore use the giant stars in our sample and the Sun (as its surface average magnetic field strength is $\sim$0.001 kG) to understand what is our magnetic field detection limit. For the Sun and for TYC 1293-2421-1, we measure a magnetic field strength of $\langle \rm B \rangle \sim \, 0.1 \,kG$, while for BD+004988 and EPIC 211304446 we measure magnetic field strengths of $\sim 0.2$ kG and  $\sim 0.3$ kG, respectively. Because in our models none of the magnetic field values of these stars converged to zero, we define our magnetic field detection limit to be $\langle \rm B_{ limit} \rangle$ = 0.31 kG, as it is the strongest magnetic field value among the sample of stars where we expected a null field. From now on, we will use this limit as a metric to define whether or not the magnetic field of a star is detected.

\section{Modeling Young Stars}
\label{sec:magnetic-models}

\subsection{Magnetic Model Verification: BP Tau}
\label{subsec:BP_Tau}
\begin{figure*}[ht!]
\epsscale{1.0}
\plotone{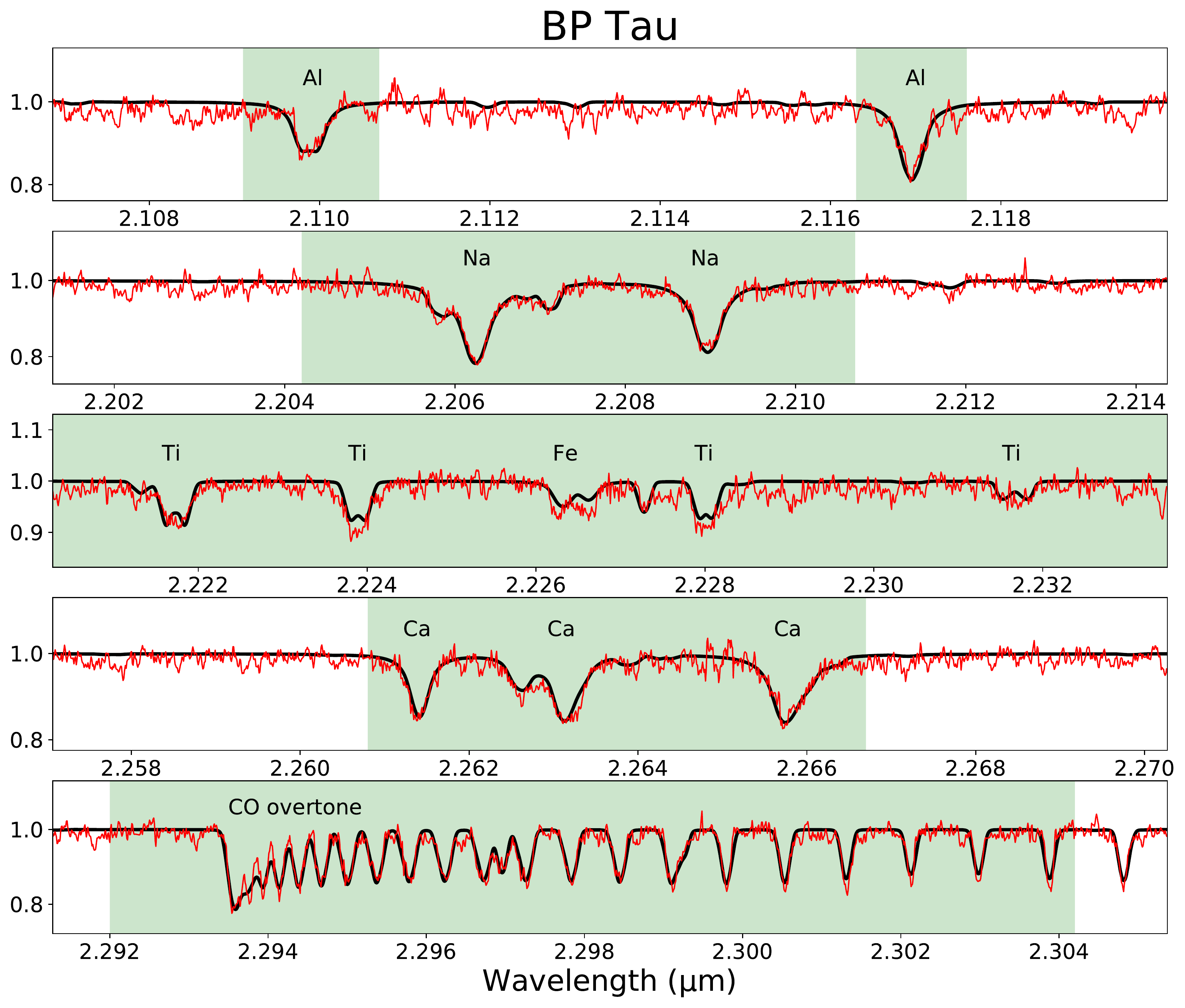}
\caption{Comparison between the observed spectrum of BP Tau (in red) and our best-fit model (in black) from MoogStokes. The six green panels show the six wavelength regions we used to fit the spectrum of the young star. \label{fig:BP_Tau_best_fit}}
\end{figure*}

As a next step, we now turn to young stars and apply the method described in Section \ref{subsect:how_we_fit_models} to the class II source BP Tau. This star was chosen because it already has a measurement of its magnetic field strength \citep{Johns-Krull1999,Johns-Krull2007}, and it thus serves as a test of our models applied to young stars. We started the modeling of this young star by fixing the metallicity at [M/H] = 0.0 and allowing the r$_K$ parameter to vary. Metallicity and IR K-band veiling parameters have a similar behavior in the spectra of stars, both making the lines appear to be stronger or weaker, and therefore metallicity and IR K-band veiling are partially degenerate. We allowed the magnetic field parameter $\langle \rm B \rangle$ to vary by setting a uniform prior distribution for the magnetic field strength between 0 and $\rm 3.5\, kG$. In our calculations we adopted a single temperature and a single surface magnetic field strength to model the surface of the young stars. Models that include two temperatures, one for the spotted and one for the nonspotted regions, and a distribution of magnetic field strengths might be a better representation of the stars \citep{Debes2013,Gully-Santiago2017}. However, including all these extra parameters in our models would make the computational time prohibitively large. To derive BP Tau's stellar parameters, we ran $80,000$ models, using all six wavelength regions defined in Section \ref{subsec:diagnostic_lines}, fitting simultaneously T$_{\rm eff}$, log(g), $\langle \rm B \rangle$, r$_{K}$, $v\sin(i)$, $v_{\rm micro}$, and the CO abundance. In Figure \ref{fig:BP_Tau_best_fit} we show the best fit magnetic model overplotted with the spectrum of BP Tau, and in Table \ref{table:YSO_stellar_params} we summarize the stellar parameters derived.

Several authors have measured the atmospheric stellar parameters of BP Tau. \cite{Schiavon1995} used atomic and molecular line depth ratios to determine a $\rm T_{\rm eff} = 4\,060\, K$, $\rm \log(g) = 4.3$ and $v \sin(i)<10 \rm \, km \, s^{-1}$. \cite{Hartigan1995} used a combination of optical spectroscopy and photometry to place BP Tau on the Hertzsprung-Russell (HR) diagram and derive $\rm T_{\rm eff} = 4\,000\rm K$ and $\rm \log(g) = 3.53$ using \cite{D'antona1994} evolutionary models. \cite{Grankin2016} measured an effective temperature of $\rm T_{\rm eff} = 4\,000\, K$ for BP Tau, using the \cite{Tokunaga2000} spectral-type scale  and then derived a surface gravity of $\rm log(g) = 3.76$ by combining \cite{Siess2000} evolutionary models with long-term optical photometry. \cite{Johns-Krull1999} used optical high spectral resolution observations and stellar synthetic spectral models to derive a $\rm T_{\rm eff} = 4\,055 \pm 112\, K$, $\rm \log(g) = 3.67 \pm 0.5$, $\rm [M/H] = 0.18 \pm 0.11$, and a $v\sin(i) = 10.2 \pm 1.8 \, \rm km \, s^{-1}$. They also used NIR spectroscopic observations and magnetic synthetic spectral models to measure the average surface magnetic field strength of BP Tau. \cite{Johns-Krull1999} used a single temperature model and a distribution of magnetic fields on the surface of the star to measure an average magnetic field strength value of $\langle \rm B \rangle = 2.6 \pm 0.3 \rm \,kG $ for BP Tau. Using a different set of observations but a similar description for the distribution of magnetic fields, \cite{Johns-Krull2007} adopted a spectral-type measurement from \cite{Gullbring1998}, a surface gravity derived from the \cite{Siess2000} models, and a stellar metallicity of [M/H] = 0.0 to measure an average magnetic field strength of $\langle \rm B \rangle = 2.17 \pm 0.3 \rm \, kG$ for BP Tau. In both cases, \cite{Johns-Krull1999} and \cite{Johns-Krull2007} adopted a macroturbulence value of $v_{\rm macro} = 2 \rm \, km \, s^{-1}$ which accounted for both the macro- and microturbulence. 

Although we have not made the same assumptions about the magnetic field distribution on the surface of BP Tau and we did not find the same atmospheric stellar parameters  as \cite{Johns-Krull1999}, the average surface magnetic field measured in this work $\langle \rm B \rangle = 2.5^{+0.15}_{-0.16} \rm \,kG$ agrees within uncertainties with the magnetic fields derived by \cite{Johns-Krull1999} and by \cite{Johns-Krull2007}. We interpret this finding in a similar way to \cite{Yang2005}, where changes in temperature of a few hundred Kelvin and up to $\sim0.5$ in gravity do not significantly impact the measurement of the average magnetic field strengths of stars, as long as the model spectrum fits nonmagnetic lines well. 

The surface gravity we derived for BP Tau ($ \rm \log(g) = 4.3 \pm 0.14$) agrees very well with the value derived by \cite{Schiavon1995}, but it is also consistent within a $1\sigma$ uncertainty with the value derived by \cite{Johns-Krull1999}. The temperature we measured for BP Tau, however, is $\sim$400 K lower than any of the optical measurements mentioned above. We cannot rule out that such difference might be explained by the differences in the fitting techniques and the assumptions made in previous studies \citep{Johns-Krull1999}. However, due to the increasing amount of evidence showing a dichotomy between optical and NIR temperature measurements \citep{Gully-Santiago2017,Guo2018,Sokal2018}, we interpret the temperature difference as due to starspots on the surface of BP Tau. In this scenario, it is reasonable to expect that the IR determination of stellar parameters provides lower temperatures than optical studies, as a large fraction of the flux received in the K band is from the cooler regions in the stellar photosphere. This means that the effective temperature of BP Tau is likely to be in the range $\rm T_{\rm eff} \sim 4\,000 \, K - 3\,600 \, K$. However, we think that neither the optical temperature of $\sim4\,000 \,\rm K$ nor the near-infrared temperature of $\sim3\,600 \,\rm K$ truly represent the effective temperature of BP Tau\footnote{In cases where stars are covered by starspots, the effective temperature is then the average of the hot and cold components weighted by the corresponding starspot filling factors (\textit{f}). $ \rm T_{eff} = (T^4_{hot}(1-\textit{f}_{cool})+T^4_{cool} \textit{f}_{cool})^{1/4}$.}. Examples of this situation have been documented in the literature \citep{Gully-Santiago2017,Sokal2018} and might become even more evident once IR spectroscopic surveys can be performed on a large number of young stars.

\subsection{Application to the Class I Source V347 Aur}
\label{subsection:V347Aur}
In the previous section, we demonstrated that our magnetic models can reproduce the magnetic field of a class II star, now we turn to the class I protostellar source V347 Aur. We emphasize that the magnetic fields of protostars are virtually unexplored, with only one measurement in the literature performed by \cite{Johns-Krull2009}.

\begin{figure*}[ht!]
\epsscale{1.0}
\plotone{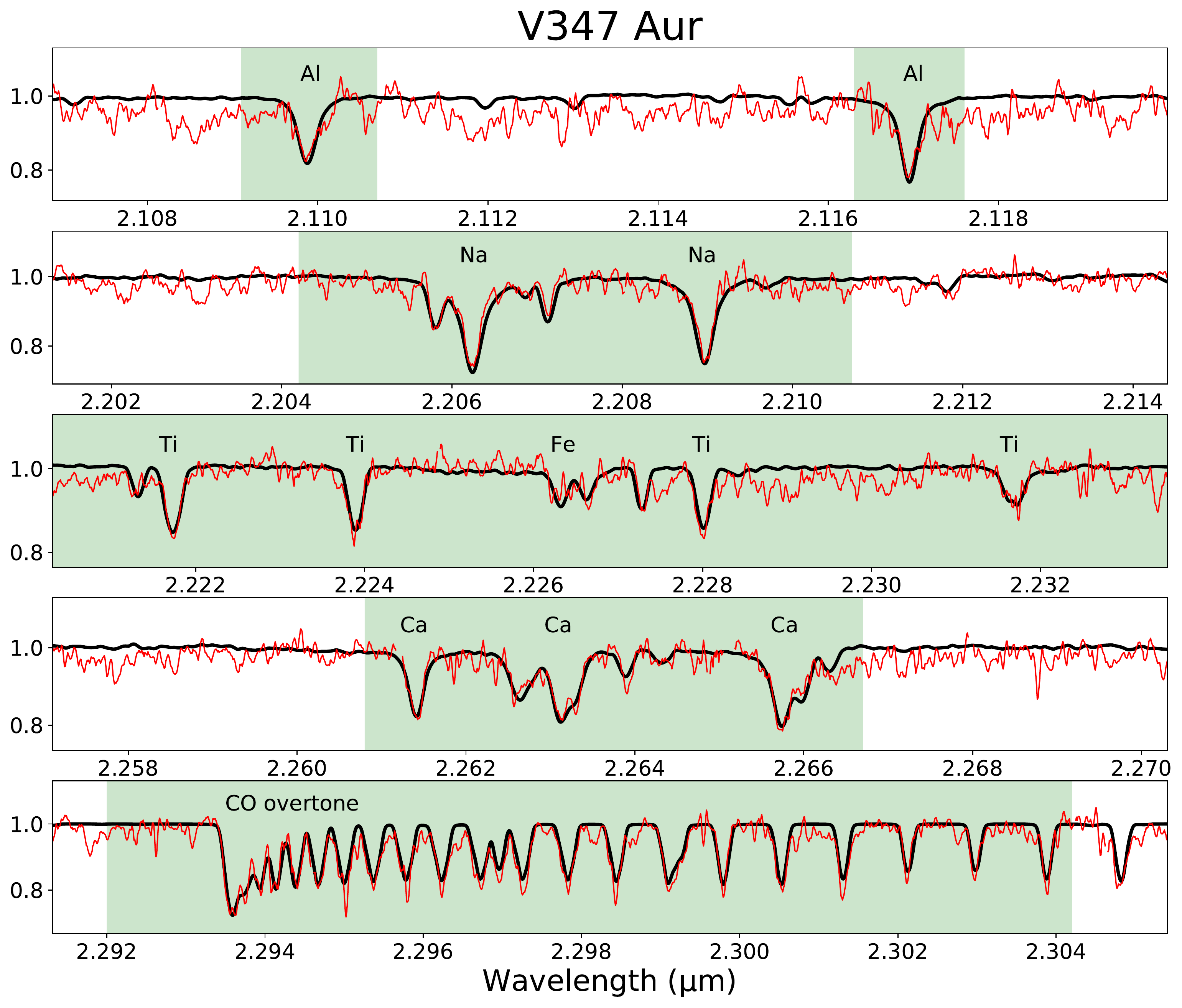}
\caption{Same as Figure \ref{fig:BP_Tau_best_fit} for the class I source V347 Aur. \label{fig:V347_Aur_best_fit}}
\end{figure*}

V347 Aur is a class I protostar in the L1438 Bok globule between Auriga, Perseus, and Camelopardalis at a distance of 208 $\pm$ 4 pc, which we obtained by inverting true \textit{Gaia} DR2 parallax \citep{Gaia2016,Gaia2018,Reipurth2008}. To measure the stellar parameters of this protostar, we followed the procedure described in the previous section, running 80\,000 magnetic models to simultaneously fit the T$_{\rm eff}$, log(g), $\langle \rm B \rangle$, r$_{K}$, $v\sin(i)$, $v_{\rm micro}$, and the CO abundance parameters. We show the best-fit model for V347 Aur in Figure \ref{fig:V347_Aur_best_fit} and in Table \ref{table:YSO_stellar_params} we summarize its best model parameters. 

We find that the best model temperature and surface gravity values for V347 Aur are $\rm T_{\rm eff} = 3\,233^{+96}_{-94} \,K$ and $\rm \log(g) = 3.25^{+0.14}_{-0.14}$, and the best average surface magnetic field is $\langle \rm B \rangle = 1.36^{+0.06}_{-0.05} \, \rm kG$. With this, we confirm that V347 Aur hosts a substantial magnetic field on its surface, as its detected magnetic field strength is much larger than our magnetic field detection limit $\langle \rm B_{limit} \rangle = 0.31 \, \rm kG$ (see Section \ref{subsection:magnetic_detection_lims}).

The temperature and gravity  measured for V347 Aur are contained within the range of stellar parameters we investigated in Section \ref{section:standard_stars}. Furthermore, the gravity we measured for the class I source lies between the gravities of giant and dwarf stars, which is expected for gravities of pre-main-sequence stars \citep{Siess2000,Baraffe,Feiden}. The temperature  and gravity  measured for V347 Aur are lower than the values derived for BP Tau. A lower temperature of a YSO on the Hayashi track means a less massive star, while a lower gravity corresponds to a less evolved star \citep{Baraffe,Feiden}. The lower gravity of the class I source V347 Aur, compared to BP Tau, is then consistent with the Lada classification of the stars, where class I sources are thought to be younger than class II sources \citep{Lada1984}. The magnetic field strength of V347 Aur has not been reported in the literature before, and it is weaker by a factor of 1.8 $\pm$ 0.1 than the magnetic field strength of BP Tau. This finding is interesting as the only other Class I source with a measured magnetic field is WL 17, which has a measured surface magnetic field strength of $\langle \rm B \rangle$ = 2.9 $\pm$ 0.43 kG \citep{Johns-Krull2009}, a value that is about twice as strong as V347 Aur's magnetic field. The range in magnetic field strength values of class I sources needs to be further explored as virtually nothing is known about magnetic fields at this early stage of stellar evolution.


Contrary to the case of BP Tau, there is only one study that measured the nonmagnetic stellar parameters of V347 Aur. \cite{Connelley} obtained a low spectral resolution (R = 1$\,$200) NIR observation of V347 Aur and fit the equivalent widths of 50 absorption lines to classify this source as a M2$^{+1}_{-1}$ star, which corresponds to a T$_{\rm eff}$ = 3\,400 -- 3\,700 K in the temperature scale of \cite{Herczeg2014}. Additionally, \cite{Connelley} classified V347 Aur as a low gravity source, due to the triangular shape of the infrared H-band spectrum. Our derived stellar parameters of T$_{\rm eff}$ = 3\,233$^{+96}_{-94}$ K and log(g) = 3.25$^{+0.14}_{-0.14}$ agree reasonably well with the values derived by \cite{Connelley}.

\cite{Doppmann} used NIRSPEC observations and non-magnetic synthetic spectral models to measure the temperatures and gravities of several class I sources. While they did not observe V347 Aur, they found three sources with effective temperature of T$_{\rm eff}$ $\sim$3\,300 K and gravities with log(g)$<$3.7, which supports our finding of a class I source with these sets of stellar parameters.

If we compare the stellar parameters we derived for V347 Aur with the ones derived for WL 17, i.e., T$_{\rm eff}$ = 3\,400 K, log(g) = 3.5, $\langle \rm B \rangle$ = 2.9 $\pm$ 0.43 kG, and $v\sin(i)$ = 11.7 $\pm$ 0.4 km s$^{-1}$ \citep{Doppmann,Johns-Krull2009}, we can infer that class I sources with similar stellar parameters can host significantly different magnetic field strengths (this is also seen in the CTTS). This can be either interpreted from the classical dipolar magnetospheric accretion theory \citep{Koenigl1991, Shu1994}, which would imply that V347 Aur and WL 17 have very different mass accretion rates or very different rotation periods (different inclination angles, given that both have similar $v\sin(i)$), but it could also mean that magnetic fields have complex morphologies and their strengths depend on stellar parameters other than rotation and accretion \citep{Donati2008,Gregory2008,Donati2009,Romanova2015,Donati2017}.

As discussed in Section \ref{subsec:BP_Tau}, the temperature we measured for V347 Aur might not correspond to the effective temperature of the star if a substantial part of V347 Aur's surface is covered by starspots. However, because there are no optical measurements of V347 Aur's temperature, we cannot confidently confirm this assumption.

\begin{deluxetable}{l|cc}
\tabletypesize{\scriptsize}
\tablecaption{Derived stellar parameters of BP Tau and V347 Aur. \label{table:YSO_stellar_params}}
\tablehead{
\colhead{YSO name} &  \colhead{BP Tau} & \colhead{V347 Aur}}
\startdata
$\rm T_{eff}$ (K) & 3\,640$^{+94}_{-92}$ & 3\,233$^{+96}_{-94}$ \\
log(g) & 4.32$^{+0.14}_{-0.14}$ & 3.25$^{+0.14}_{-0.14}$ \\
$\langle \rm B \rangle$ (kG) & 2.5$^{+0.15}_{-0.16}$ & 1.36$^{+0.06}_{-0.05}$ \\
$v\sin(i)$ (km s$^{-1}$)& 11.4$^{+0.25}_{-0.55}$ &  11.7$^{+0.16}_{-0.24}$ \\
$v_{\rm micro}$ (km s$^{-1}$) & 2.04$^{+0.08}_{-0.03}$ & 1.38$^{+0.15}_{-0.16}$ \\
r$_{\rm K}$ & 1.08$^{+0.02}_{-0.02}$ & 0.97$^{+0.02}_{-0.02}$ \\
\enddata
\end{deluxetable}

\section{Discussion}
\label{sec:discussion}

\subsection{Stellar Evolutionary Models}
Masses and ages of stars are two fundamental astrophysical parameters that are often poorly constrained. The mass of a star can be precisely measured through spectroscopy and direct imaging if the object of interest is part of a binary or a higher order system \citep{Stassun2014,David2016,Rizzuto2016}. Alternatively, masses of young stellar objects with large and bright disks can be inferred by observing (typically in the submillimeter) the motion pattern of the disk around the star, and modeling it with a Keplerian velocity profile \citep[e.g.,][]{Huelamo2015,Simon2017}. Stellar ages, on the other hand, are even harder to constrain, as only indirect measurements such as lithium abundances, rotation rates, or membership of associations can be established. In the case of very young stars, lithium abundances and rotation rates do not provide useful constraints on the stellar ages \citep{Soderblom}. Membership of young associations provide only an average age for a star-forming region (SFR), lacking the details of the age gradients recently measured in some SFRs \citep{Beccari2017,Getman2018}. In this section, we combine stellar evolutionary models with the derived stellar parameters of the YSOs to demonstrate the pre-main-sequence status of our sources. Although we do not determine a firm age and mass for our stars, we illustrate the impact that starspots can produce in the derived masses and ages of young stars.

\begin{figure*}[ht!]
\epsscale{1.0}
\plotone{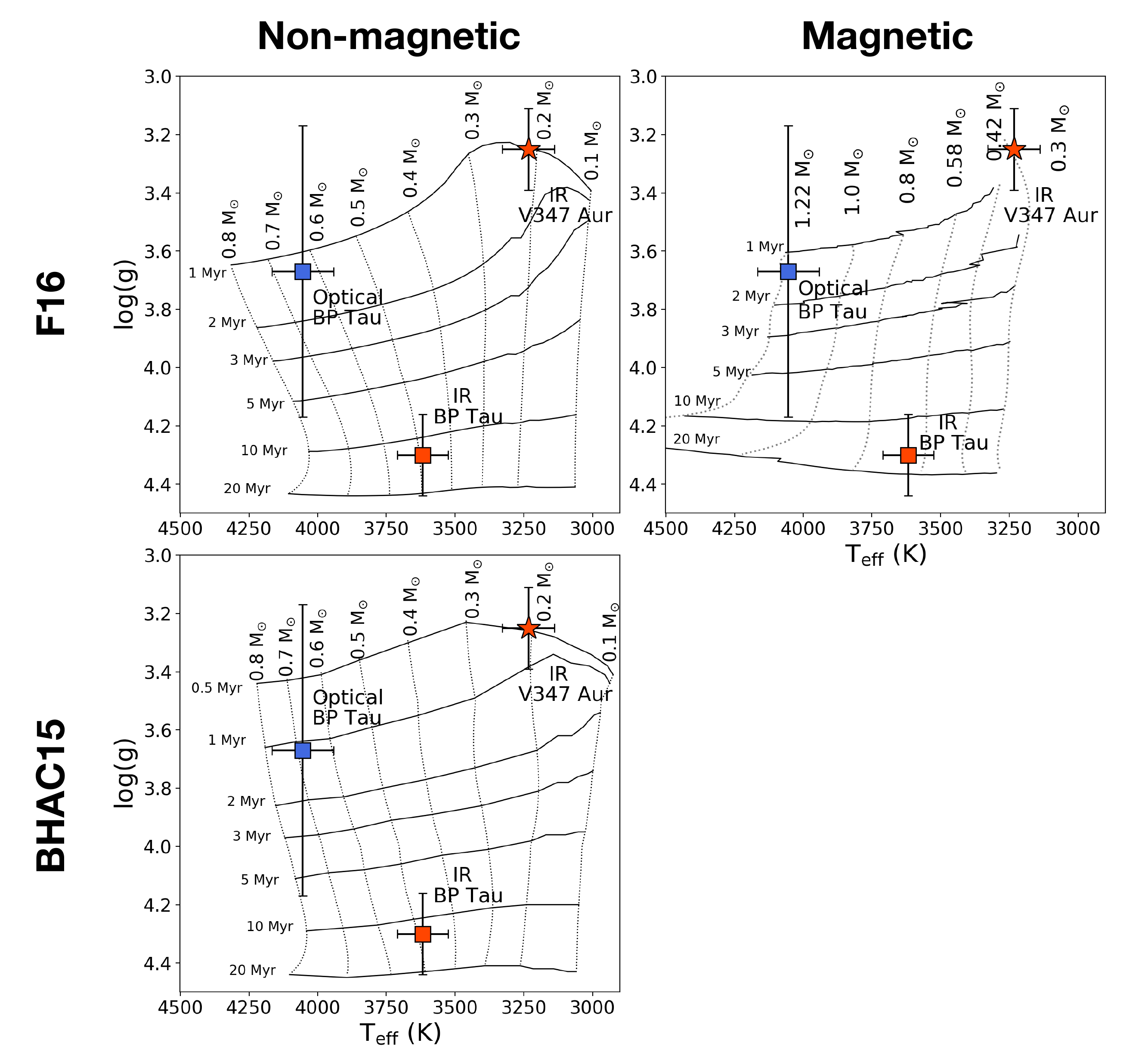}
\caption{Comparing the position of BP Tau and the class I source V347 Aur in an HR diagram with different stellar evolutionary models. The first row shows the evolutionary models from \cite{Feiden}, while the second row shows the model from \cite{Baraffe}. The positions of V347 Aur and BP Tau are plotted using the best values we derived in Section \ref{sec:magnetic-models} (red symbols). The stellar parameters for BP Tau derived from optical observations are plotted as a blue square \citep{Johns-Krull1999}. The difference between the optical and near-infrared measurements of BP Tau shows the effects of starspots on the surface of young stars. 
\label{fig:HRD_Tracks}}
\end{figure*}

In Figure \ref{fig:HRD_Tracks}, we plot the temperature and gravity we measured for BP Tau and V347 Aur (red symbols) superimposed on the \cite{Baraffe} (BHAC15) and \cite{Feiden} (F16) evolutionary models. The nonmagnetic evolutionary models of F16 and BHAC15 provide very similar values of masses and ages for both young stars. The differences at the earliest ages and lower masses correspond to differences in initial conditions chosen for each model, such as the treatment of deuterium burning in the interior of the stars. The F16 magnetic models, on the other hand, predict higher masses and also slightly older ages for a given effective temperature and surface gravity than the nonmagnetic models. This difference is produced because models that include magnetic fields in their calculations partially inhibit convection in the stellar interiors (starspots are examples of such convection inhibition). When convection is inhibited, a larger fraction of stellar energy gets trapped inside the star, decelerating the contraction process. Larger and cooler stars would then appear to be younger and less massive, which explain the above-mentioned characteristics \citep{Feiden}.

In addition to our temperature and gravity measurements, we plotted the temperature and gravity results for BP Tau from \cite[][blue symbol]{Johns-Krull1999}. We emphasize that \cite{Johns-Krull1999} used high-resolution optical spectroscopic data to derive the stellar parameters for BP Tau; therefore, their results are more likely to represent a nonspotted region on the surface of the young star. We interpret our near-infrared results as being more affected by the spotted regions on the surface of the star, which explain the lower temperature measured; however, as mentioned in Section \ref{sec:magnetic-models} temperature differences due to variations in the modeling techniques cannot be completely ruled out.

When we use stellar evolutionary models, the difference between the measured optical and NIR temperatures for BP Tau translates into a large uncertainty in its derived mass. For example, if we assume a gravity of $\rm \log(g) = 4.15$ for BP Tau, to be consistent with \cite{Johns-Krull1999} and our results, and use the nonmagnetic evolutionary models of BHAC15, we derive a stellar mass that ranges between $\sim 0.5\, \rm M_{\odot}$ and $\sim 0.8\, \rm M_{\odot}$. If, on the other hand, we consider the F16 magnetic evolutionary models, then the derived mass for BP Tau ranges between $\sim 0.6\, \rm M_{\odot}$ and $\sim 1.1\, \rm M_{\odot}$. As can be seen from Fig \ref{fig:HRD_Tracks}, the optical measurement provides the high-end mass estimates of BP Tau, while NIR temperatures provide the low-end mass estimate of it. This is true for low-mass young stars as they contract almost isothermally during the first few megayears of their evolution.

Unfortunately, current pre-main-sequence stellar evolutionary models do not deal with heavily spotted stars, and it is therefore difficult to retrieve a precise mass for a star with temperature measurements in both regions, the spotted and nonspotted regions. Notwithstanding, it is possible to assign an effective temperature to a young star, if the covering factor of the spotted and nonspotted regions along with the temperature of both regions iw known; see, for example, \cite{Gully-Santiago2017}. For the moment, however, we acknowledge the problem of deriving masses and ages in young stars using a single temperature component model and offer Figure \ref{fig:HRD_Tracks} as a good example of this problem.

\subsection{Nonmagnetic Models when Estimating Stellar Parameters of YSOs}
We have investigated how the atmospheric stellar parameters change when we use nonmagnetic models to fit the spectrum of YSOs. These types of studies have been performed in the past and the most common outcome is that stellar temperatures change by a few hundred Kelvin, surface gravities vary by up to 0.5 dex, and rotational velocities differ by more than $2 \rm \,km \,s^{-1}$ \citep{Doppmann2003,Sokal2018}. To set new tests to this long-lasting assumption, we modeled the spectrum of our two YSOs fixing the magnetic field strength to $\langle \rm B \rangle = 0 \rm \, kG$, and allowing the T$_{\rm eff}$, log(g), r$_K$, $v\sin(i)$, $v_{\rm micro}$, and CO abundance to vary. In Table \ref{tab:non-magnetic-YSO-models} we summarize the results from this experiment. 

We found that the temperatures and gravities derived by the nonmagnetic models are consistent with the results from the magnetic models. However, the derived rotational velocity and the microturbulence values of the nonmagnetic models are larger than results from the magnetic models. The rotational velocity and the microturbulence values of BP Tau increased by a factor of 1.26 and 1.04, while the same parameters increased by 1.17 and 2.1 for V347 Aur. We assume that the increase of the rotational velocity and microturbulence values of the stars partially compensate for the lack of magnetic broadening in the spectral lines (especially in the most magnetically sensitive lines). Modifying the rotation and the turbulence of the star, however, yields a worse fit to the CO lines, and still an inadequate fitting to the width of the Ti lines. Therefore, contrary to other studies where the exclusion of magnetic fields affect the derived temperature and gravity of the YSOs, the ``nonmagnetic version'' of our method mainly affects the rotation rate of the star, although providing a worse overall fit to the data.

\begin{table*}[!hbt]
   \centering
   \caption{Comparison between Magnetic and Nonmagnetic Synthetic Spectral Models Applied to our YSOs} \label{tab:non-magnetic-YSO-models}
   \begin{tabular}{c | c c | c c }
      \toprule
       & \multicolumn{2}{c}{BP Tau} & \multicolumn{2}{c}{ V347 Aur} \\
      Model type & Magnetic & Nonmagnetic & Magnetic & Nonmagnetic \\
       \hline
      T$_{\rm eff}$ (K) & 3\,640$^{+94}_{-92}$ & 3\,600$^{+92}_{-92}$ & 3\,233$^{+96}_{-94}$ & 3\,203$^{+92}_{-92}$  \\
      log(g) & 4.32$^{+0.14}_{-0.14}$ & 4.34$^{+0.14}_{-0.14}$ &  3.25$^{+0.18}_{-0.18}$ & 3.25$^{+0.18}_{-0.18}$ \\
      $\langle \rm \overrightarrow{\rm B} \rangle$ (kG) & 2.5$^{+0.15}_{-0.16}$ & - &  1.36$^{+0.06}_{-0.05}$ & -  \\
      $v\sin(i)$ (km s$^{-1}$)& 11.4$^{+0.25}_{-0.55}$ & 16.6$^{+0.47}_{-0.38}$ &   11.7$^{+0.16}_{-0.24}$ & 13.5$^{+0.12}_{-0.35}$ \\ 
      $v_{\rm micro}$ (km s$^{-1}$) & 2.04$^{+0.08}_{-0.03}$ & 2.14$^{+0.24}_{-0.11}$ & 1.38$^{+0.15}_{-0.16}$ & 2.96$^{+0.15}_{-0.14}$ \\
      r$_{\rm K}$ & 1.08$^{+0.02}_{-0.02}$ & 0.87$^{+0.02}_{-0.02}$ &  0.97$^{+0.02}_{-0.02}$ & 1.05$^{+0.02}_{-0.02}$ \\
      \bottomrule
   \end{tabular}
\end{table*}




\section{Conclusion} \label{sec:Summary}
In this study, we have developed a method to calculate the atmospheric stellar parameters of young stars. We have tested the method using iSHELL high-resolution NIR observations and the magnetic radiative transfer code MoogStokes. A summary of our findings are as follows:
\begin{itemize}
  \item We recalculated the line transition parameters of 26 atomic lines in the K band,  to enhance the prediction power of our method. We did this by comparing iSHELL solar observations (reflected light from the asteroid Ceres) to spectral synthesis models.
  \item  To gauge the true uncertainties in the stellar parameters derived from our method, we modeled a set of nine main- and post-main-sequence stars. We obtained a temperature uncertainty of $\sigma_{Teff} = 91 \rm \, K$, a gravity uncertainty of $\sigma_{\rm \log(g)} = 0.14$, and a magnetic field detection limit of $\rm \langle \rm B_{limit} \rangle = 0.31 \, \rm kG$. 
  \item We applied the synthetic spectrum method to the class II source BP Tau and found that our derived gravity $\rm \log(g)=4.32^{+0.14}_{-0.14}$ and magnetic field strength $\langle \rm B \rangle = 2.5^{+0.15}_{-0.16} \, \rm kG$ agree within uncertainties with previous studies. Our derived temperature, however, is lower than any of the temperatures derived from optical studies in the literature. We interpret our lower temperature measurement as due to the effect of starspots on the surface of BP Tau, which more strongly affect the IR observations than optical observations. 
  
  \item We applied the same modeling method to the class I protostellar source V347 Aur, and we measured for the first time its surface gravity $\rm \log(g)=3.25^{+0.14}_{-0.14}$, projected rotational velocity $ v \rm \sin(i)=11.7^{+0.16}_{-0.24}$, and magnetic field strength $\langle \rm B \rangle = 1.36^{+0.06}_{-0.05} \, \rm kG$. We highlight the importance of measuring magnetic fields in class I sources, for which only one other case has such a measurement. We are currently carrying out a large infrared spectroscopic survey of class I sources to understand how magnetic fields affect protostars at early stages of stellar evolution.
  \item Finally, we have combined the measured stellar parameters of BP Tau and V347 Aur with pre-main-sequence stellar evolutionary models to demonstrate their pre-main-sequence status and also to illustrate the difference between the masses and ages derived from optical and NIR observations. Although we do not provide definite ages for any of the YSOs, we can confidently state that BP Tau is older than V347 Aur, which is  consistent with the Lada classification of both sources.
\end{itemize}

\acknowledgements
\section*{Acknowledgements}
We would like to thank Curt Dodds for his help installing the MoogStokes code on the IfA servers, to Sam Grunblatt for sharing a list of red giant stars with good asteroseismic measurements, and to Michael Lum for providing a script that helps with the interpolation between MARCS stellar atmospheric models. We also thank the referee for the insightful review which helped to improve this paper. We acknowledge the support of the NASA Infrared Telescope Facility, which is operated by the University of Hawaii under contract 80HQTR19D0030 with the National Aeronautics and Space Administration, and we are grateful for the professional assistance in obtaining the observations from Dave Griep, Eric Volquardsen, Brian Cabreira, Tony Matulonis, and Greg Osterman. This research has made use of the SIMBAD database, operated at CDS, Strasbourg, France, and NASA’s Astrophysics Data System. This work has made use of the VALD database, operated at Uppsala University, the Institute of Astronomy RAS in Moscow, and the University of Vienna. This work has also made use of data from the European Space Agency (ESA) mission
{\it Gaia} (\url{https://www.cosmos.esa.int/gaia}), processed by the {\it Gaia}
Data Processing and Analysis Consortium (DPAC,
\url{https://www.cosmos.esa.int/web/gaia/dpac/consortium}). Funding for the DPAC
has been provided by national institutions, in particular the institutions
participating in the {\it Gaia} Multilateral Agreement.

%

\vspace{5mm}
\facilities{IRTF}


\software{Astropy \citep{Astropy2018},  
          MOOG \citep{Sneden}, 
          MoogStokes \citep{Deen}, 
          emcee \citep{Foreman-Mackey}, 
          Spextools \citep{Cushing2004}, 
          xtellcor \citep{vacca2003}
          }



\appendix
\section{iSHELL Spectral Profile} \label{sec:appendix_iSHELL_PSF}
\subsection{Empirical Measurement}
\subsubsection{Arc-lamp Data Acquisition and Data Reduction}
Detailed line profiles studies such as those presented in this work require accurate instrumental profile characterization. These instrumental profiles can be measured using extremely narrow emission or absorption lines. To characterize the iSHELL spectral profile, we used the narrow emission lines from the Th--Ar lamp built into iSHELL. On 2018 February 2, we collected iSHELL calibration data (arcs and flat files) in the K2 mode using the 0\farcs375 and 0\farcs75 slit widths. We obtained nine standard calibration observations for each slit width, where each calibration observation corresponds to five 7 s exposure flat files and two 14 s thorium--argon lamp files (lamp on and off). 
We reduced the calibration data using \texttt{Spextool v5.0.2} as indicated in the Spextool manual \citep{Cushing2004}, with the only exception that we considered our arc-lamp calibration files as raw astronomical data. Using the \texttt{xspextool} task, we first created the normalized flat-field images and wavelength calibration files. Then, we extracted the spectra of the arc-lamps in the extended source configuration (as the arc-lamp light completely fills the detector's slit width entrance). To increase the S/N of the arc-lamp file, we used \texttt{xcombspec} to combine the nine multiorder lamp spectra and \texttt{xmergeorders} to merge multiorder spectra into a single continuous spectrum. Finally, we used \texttt{xcleanspec} to eliminate any deviant or negative pixel caused by imperfections in the infrared array.

\subsubsection{Intramode Wavelength Dependence} \label{subsection:intra-model-wave-dep}
\begin{figure}[ht!]
\epsscale{1.0}
\plotone{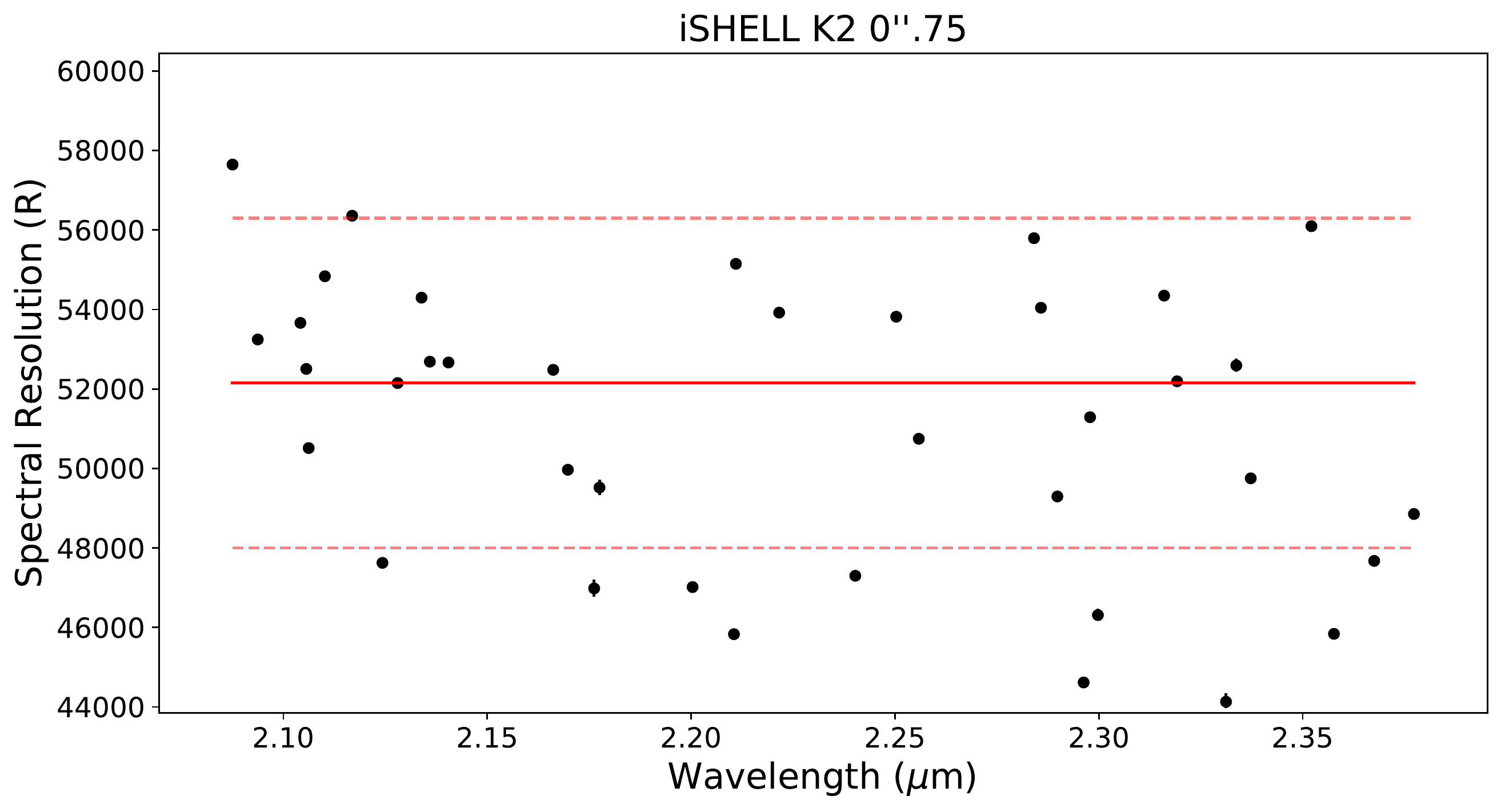}
\caption{Spectral resolution as a function of wavelength measured from individual arc-lamp lines in the K2 mode of iSHELL. We fit a Gaussian model to each arc line to determine the wavelength position and spectral resolution $R$. The red solid line corresponds to the median of the spectral resolution and the dashed lines to the 1$\sigma$ dispersion. \label{fig:iSHELL_spectral_dependence}}
\end{figure}

To test if iSHELL's spectral resolution changes with wavelength within the K2 mode, we measured the FWHM (full width at half maximum) and wavelength position of individual thorium--argon lines from 2.09 \micron \, to 2.36 \micron. First, we visually selected single (not blended), noncontaminated, and nonsaturated lines in the full extent of the K2 mode. Then, we fit each emission line with a Gaussian plus a linear model to obtain a measurement of the centroid (central wavelength of the line), the peak of the line, the baseline (continuum emission), and the FWHM of each selected arc-lamp line. In Figure \ref{fig:iSHELL_spectral_dependence} we plot the best-fit (Gaussian model) spectral resolution $R$ as a function of wavelength for the 39 lines we identified in the K2 mode. We obtained a median spectral resolution of R = $52\,000$ and a 1$\sigma$ dispersion of $\rm \Delta R \:= \: 4\,000$ for the K2 mode 0\farcs75 slit width. Although we noticed a small spectral resolution dependency with wavelength on the K2 mode 0\farcs75 slit width, the measured change in the spectral resolution within the whole K2 mode ($\rm \Delta R \:= \: 4\,000$) is comparable to the spectral resolution scatter of the individual lines. Additionally, when we performed the same analysis using the 0\farcs375 slit width, we did not see any spectral resolution trend with wavelength. We list the measured R for each slit width mode in Table \ref{table:iSHELL_analytical_PSF} calculated from the empirical point-spread function (Section \ref{subsec:line-stacked-profile}).

\subsubsection{Line Stacked Spectral Profile} \label{subsec:line-stacked-profile}
\begin{figure*}[ht!]
\epsscale{1.1}
\plotone{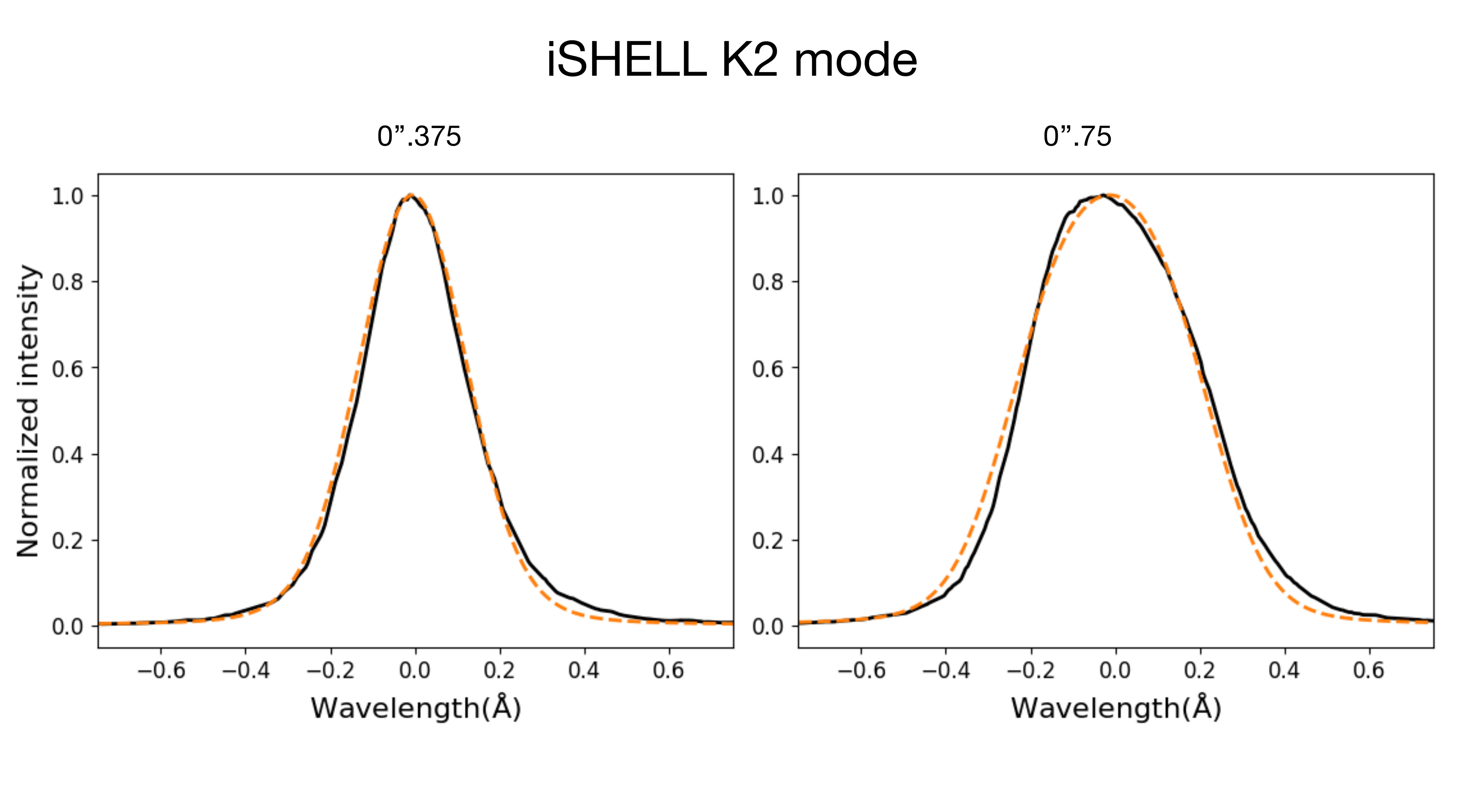}
\caption{The iSHELL spectral profile in the K2 mode 0\farcs375 and the 0\farcs75 slit widths. The left (right) box shows the measured 0\farcs375 (0\farcs75) spectral profile in black solid line. The dashed orange line corresponds to the best fit with a Voigt (Voigt$\ast$BoxCar) profile. The spectral resolution and its associated uncertainty are shown in Table \ref{table:iSHELL_analytical_PSF}.  \label{fig:iSHEL_K2_PSF}}
\end{figure*}

To obtain a high S/N and well-sampled iSHELL spectral profile, we stacked the spectral profiles of the 39 individual thorium--argon lines in the K2 mode. We first used the centroid positions, continuum emission level, and normalization constant values derived in Appendix \ref{subsection:intra-model-wave-dep} to move the individual arc-lamp lines to a common wavelength position. We then subtracted the continuum level and normalized the peak value of the lines to unity. Afterward, we took the median of the spectral profiles to obtain a stacked line profile that corresponds to the iSHELL spectral profile in the K2 mode (See Figure \ref{fig:iSHEL_K2_PSF}). To calculate the spectral resolution of the stacked profile, we measured the FWHM in angstroms and computed the spectral resolution R as the average wavelength of the K2 mode (22\,335 \AA) divided by the FWHM (See column 2 of Table \ref{table:iSHELL_analytical_PSF}). Finally, we calculated the spectral resolution uncertainty by adding in quadrature the uncertainty obtained from stacking the individual line profiles (standard deviation of the stacked profile) and the uncertainty produced by wavelength shifts of the centroid fitting.

\subsection{Analytical Spectral Profile Prescription}
To obtain an analytical prescription of the iSHELL spectral profile that can be widely used by the scientific community, we modeled the empirical iSHELL K2 spectral profile (see Appendix \ref{subsec:line-stacked-profile}) with a Gaussian, a Lorentz, a BoxCar function, and every possible combination (convolution) of the mentioned profiles. To find which profile better fits  the measured iSHELL K2 spectral profile, we implement a $\chi^2$ test with $n+1$ degrees of freedom, where $n$ is the number of functional forms used in the convolution, for instance, a simple Gaussian fit has two free parameters (the wavelength position and the Gaussian FWHM) while a Voigt profile convolved with a BoxCar function has four free parameters (three FWHM parameters and one wavelength position parameter). We found that the best fit for the 0\farcs375 slit width corresponds to a Voigt profile, while the best fit for the 0\farcs75 slit width corresponds to the convolution of a Voigt and a BoxCar profile. In Table \ref{table:iSHELL_analytical_PSF} we summarize the measured spectral resolution of the iSHELL K2 mode and the analytical approximations we use to characterize both profiles.

\begin{deluxetable}{lcccccc}
\tabletypesize{\scriptsize}
\tablecaption{iSHELL spectral response summary. \label{table:iSHELL_analytical_PSF}}
\tablewidth{0pt}
\tablecolumns{7}
\tablehead{
\colhead{Slit} &  \colhead{Spectral} & \colhead{Functional} & \colhead{Gaussian} & \colhead{Lorentzian} & \colhead{BoxCar} & \colhead{Wavelength} \\
\colhead{Width} & \colhead{Resolution (R)} & \colhead{Form} & \colhead{FWHM (\AA)} & \colhead{FWHM (\AA)} & \colhead{FWHM (\AA)} & \colhead{Shift (\AA)}}
\colnumbers
\startdata
0\farcs75 & $47\,553^{+3\,793}_{-3\,175}$ & Voigt $\ast$ BoxCar & 0.246 & 0.04 & 0.449 & -0.015 \\
0\farcs375 & $78\,751^{+7\,226}_{-6\,268}$ & Voigt & 0.28 & 0.04 & - & 0.0 \\
\hline
\enddata
\tablecomments{The $\ast$ symbol represents convolution.}
\end{deluxetable}

\section{Modification of the VALD3 Line Transition Parameters} \label{subsec:linelist-modification}
\begin{figure*}[ht!]
\epsscale{1.0}
\plotone{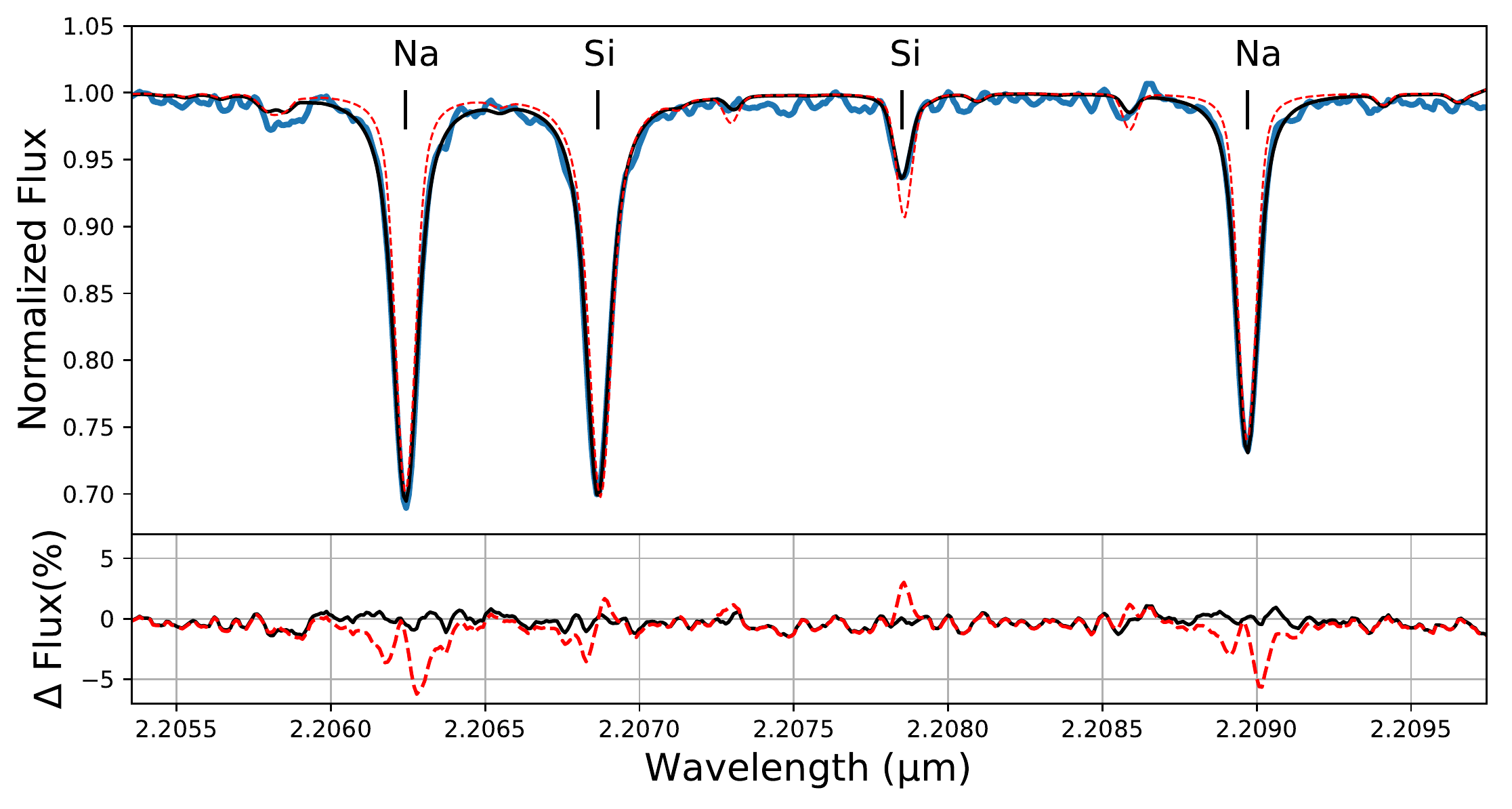}
\caption{Upper panel: comparison between the iSHELL solar observations (blue thick line), the MoogStokes model with the default VALD3 line transition parameters (dashed red line), and the MoogStokes model after we modified the line transition parameters (solid black line). The modifications to the VALD3 line list parameters significantly improved the match between the model and the observations. Lower panel: in black, residuals of the MoogStokes model with modified line list parameters. In red, residuals of the MoogStokes model with default VALD3 database line parameters. At the position of the spectral lines, the residual flux is less than 2\% and it is comparable to the noise of the line-free region. \label{fig:Solar_spec_Reg1}}
\end{figure*}

\begin{figure*}[ht!]
\epsscale{1.0}
\plotone{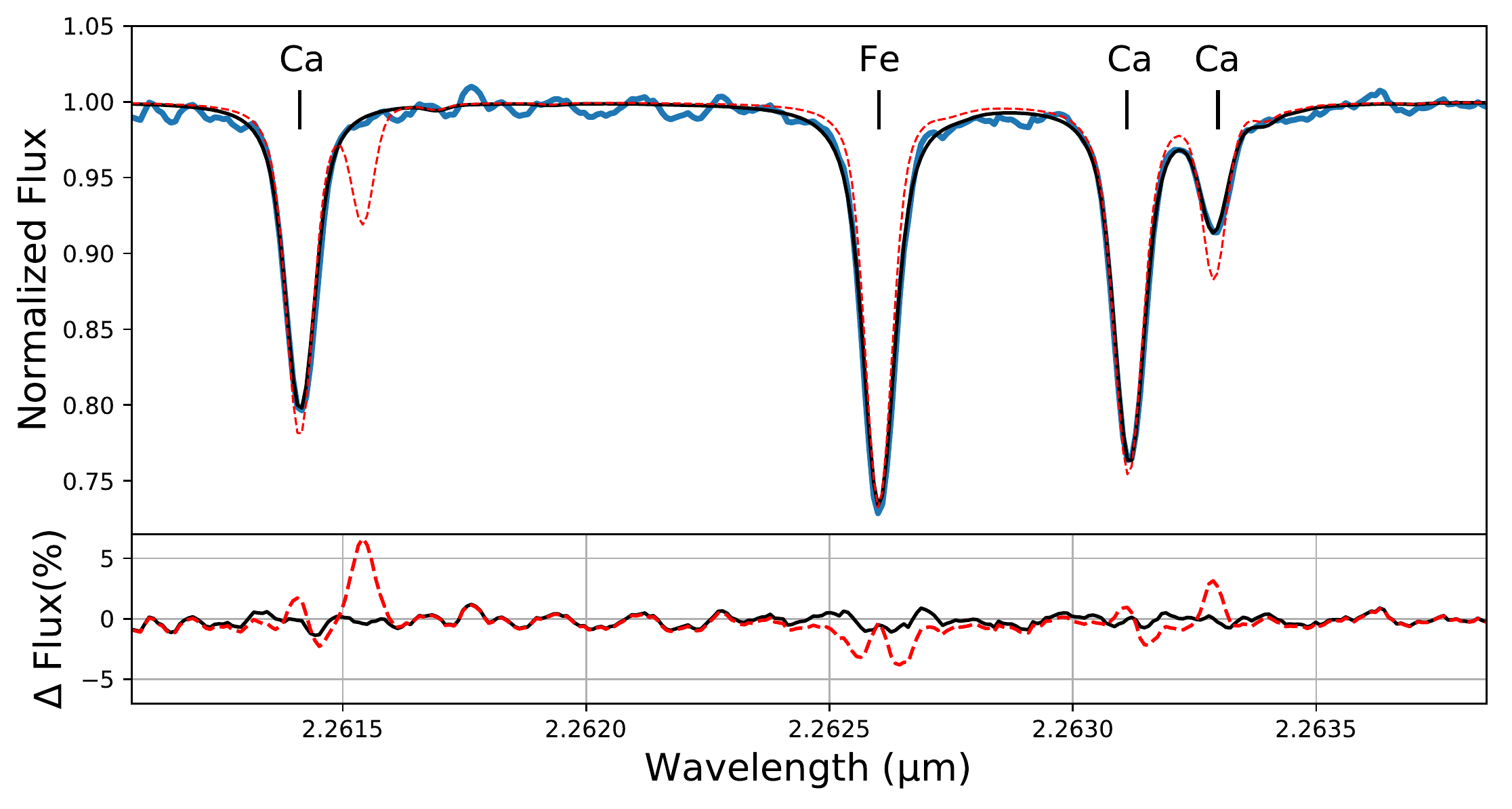}
\caption{ Same as Figure \ref{fig:Solar_spec_Reg1} for the Ca I and Fe I lines. \label{fig:Solar_spec_Reg4}}
\end{figure*}

Line transition parameters strongly affect the shape and strength of stellar spectral lines. Therefore, choosing appropriate values of these parameters is essential for a correct spectroscopic modeling study. In this section, we aim to assess the quality of the log(gf) and van der Waals constants values we use by comparing iSHELL solar observations against the synthetic solar spectrum generated with default VALD3 line transition parameters \citep{Ryabchikova2015}. We first computed a synthetic solar nonmagnetic spectrum using MoogStokes with standard solar parameters $\rm T_{eff}$ = 5\,778 K, log(g) = 4.43, metallicity [M/H] = 0.0, $v_{\rm micro}$ = 1.0 km s$^{-1}$, and $v\sin(i)$ = 2.0 km s$^{-1}$. Initially, we adopted an isotropic Gaussian macroturbulence velocity of $v_{\rm macro} = 2 \rm \,km \, s^{-1}$ \citep{Steffen2013}. However, due to uncertainties in the correct treatment of the macroturbulence velocities in the Sun \citep{Takeda2017} and the small effects ($<$1$\%$) it produces in our synthetic spectra, we decided to not include this extra parameter in our modeling. We then convolved the MoogStokes output spectrum with the K2 0\farcs75 slit width spectral profile and resampled it in wavelength space to match the observations. Figures \ref{fig:Solar_spec_Reg1} and \ref{fig:Solar_spec_Reg4} show a comparison between the observed solar spectrum and the MoogStokes model with the default line transition parameters from the VALD3 database.

The visible mismatch between the observations and the models with standard VALD3 values could be caused by, but are not limited to, inaccurate log(gf) and VdW constant values in the database, by oversimplifications made on the stellar atmospheric models (such as the LTE assumption), by imperfections in the radiative transfer code, by an imperfect characterization of our instrument’s spectral profile, or more likely by a combination of the effects mentioned above. These inaccuracies prevent us from generating perfect spectral synthesis models and therefore hamper our ability to recover precise stellar parameters. In this study, however, as in many others \citep{Shetrone2015,Andreasen} we assumed that the main source of discrepancy between the models and the data comes from the line transition parameters in the VALD3 database. For this reason, we have adjusted the log(gf) and VdW values of the spectral lines listed in Section \ref{subsec:diagnostic_lines}, until the computed synthetic spectrum matched the solar observations. However, we caution the reader that by modifying the log(gf) and VdW constant values, we could be hiding defects associated with the stellar atmospheric models, the radiative transfer code, or even the measured instrument spectral profile.
The solid black lines of Figures \ref{fig:Solar_spec_Reg1} and \ref{fig:Solar_spec_Reg4} show the solar model synthesized with the adjusted line transition parameters. The lower panel of each figure shows the residual between the solar observation and the MoogStokes model with the modified line transition parameters. In total, we modified 26 individual line transition parameters from nine different elements in the wavelength range 2.1098--2.2964 \micron. We summarize the modified line transition parameters in Table \ref{table:Line_params}.

\begin{deluxetable}{lcccccc}
\tabletypesize{\scriptsize}
\tablecaption{VALD3 Line Parameters Modified in This Work. \label{table:Line_params}}
\tablewidth{0pt}
\tablecolumns{7}
\tablehead{ \colhead{} & \colhead{} &\colhead{Default VALD3 parameters} & \colhead{} & \colhead{} & \colhead{Modified VALD3 parameters} & \colhead{}\\
\colhead{Element} &\colhead{Wavelength (\AA)} & \colhead{log(gf)} & \colhead{Waals} & \colhead{Wavelength (\AA)} & \colhead{log(gf)} & \colhead{Waals}}
\startdata
 Al I & 21098.787 & -0.309 & 0.000 & 21098.790 & -0.130  &  -7.220\\
 Fe I & 21101.160 & -0.721 & -7.340 & 21101.100 & -0.761  &  -7.140 \\
 \hline
 Fe I & 21167.811 & -0.340 & -7.320 & 21167.751 & -0.600  &  -7.320 \\
 Al I & 21169.532 & -0.009 & 0.000 & 21169.532 & 0.409  &  -7.550 \\
\hline
 Sc I & 22058.159 & -0.838 & -7.820 & 22058.059  &        -1.138  &  -7.220\\
 Na I & 22062.420 & 0.287 & 0.000 & 22062.420  &        0.407  &  -7.250\\
 Si I  & 22068.732 & 0.538 & -7.330 & 22068.655  &        0.590  &  -7.320\\
 Si I  & 22078.574 & -0.945 & -7.330  & 22078.504  &       -1.085  &  -6.930\\
 Na I  & 22089.690 & -0.013 & 0.000 & 22089.690  &       -0.000  &  -7.070\\
 V I  & 22097.543 & -0.754 & -7.800 & 22097.543  &       -1.254  &  -7.800\\
\hline
 Ti I  & 22238.926 & -1.690 & -7.790 & 22238.926   &      -1.950  &  -7.790\\
 Fe I  & 22263.182 & -0.871 & -7.540 & 22263.185   &      -0.751  &  -7.070\\
 Fe I  & 22266.255 & -0.941 & -7.540 & 22266.260   &      -0.981  &  -6.950\\
 Ti I  & 22280.101 & -1.800 & -7.790 & 22280.101   &      -2.00  &  -7.790\\
 Ti I  & 22316.706 & -2.07 & -7.790 & 22316.706   &      -2.300  &  -7.790\\
\hline
 Ca I & 22614.115 & 0.516 & -7.330  & 22614.125   &      0.426  &  -7.050\\
 Fe I\tablenotemark{a} & 22615.409 & -1.692 & -7.500  & -   &      -  &  -\\
 Fe I & 22626.011 & -0.479 & -7.540  & 22626.016   &      -0.389  & -7.150\\
 Ti I & 22627.394 & -2.740 & -7.790  & 22627.394   &      -2.830  & -6.590\\
 Ca I & 22631.137 & 0.687 & -7.330 & 22631.160   &     0.637   & -7.180 \\
 Ca I & 22632.899 & -0.216 & -7.330 & 22632.890   &      -0.346  &  -6.700\\
 S I & 22650.236 & -0.340 & -7.620  & 22650.140   &      -0.350  &  -6.920\\
 Ca I & 22657.359 & 0.847 & -7.330 & 22657.409   &       0.707  &  -6.950\\
 Ca I & 22659.762 & -0.216 & -7.330 & 22659.850   &      -0.276  &  -6.920\\
 \hline
  Si I & 22940.831 & -1.382 & -7.480 & 22940.831   &      -1.682  &  -7.480\\
  Si I & 22964.269 & -0.908 & -6.940 & 22964.269   &      -1.508  &  -6.940\\
\enddata
\tablenotetext{a}{Because this line did not appear in the solar spectrum, we removed it from the list. See Figure \ref{fig:Solar_spec_Reg4}.}
\end{deluxetable}

\end{document}